\documentclass[11pt]{article}
\usepackage{amsmath,amsbsy,amsfonts,amssymb,graphics,epsfig,float,mathrsfs,amsthm,braket}
\usepackage{graphicx}
\usepackage{enumerate}
\usepackage{derivative}
\usepackage[normalem]{ulem}

\usepackage{graphicx}
\usepackage{bm}
\usepackage{enumerate} 
\usepackage{derivative}
\usepackage{mathtools}
\usepackage{xparse}
\usepackage{tikz}
\usetikzlibrary[shadings]
\usetikzlibrary{shapes}
\usetikzlibrary{snakes}
\usetikzlibrary {arrows.meta}
\usetikzlibrary{math} 
\usetikzlibrary{3d}

\usepackage[colorlinks=true, allcolors=blue]{hyperref}

\usepackage{graphicx}
\usepackage{dcolumn}
\usepackage{bm}

\renewcommand{\sout}[1]{}

\usepackage{ifsym}
\usepackage{psfrag}
\usepackage{color}
\usepackage{caption}
\usepackage{subcaption}
\usepackage{multirow}
\usepackage{float}
\usepackage[normalem]{ulem}

\newcommand \beq{\begin{equation}}
\newcommand \eeq{\end{equation}}

\newcommand{\upd}{\mathrm{d}}
\newcommand{\gr}{\overset{\circ}{\mathbf{g}}}
\newcommand{\gi}{\bar{\mathbf{g}}}
\newcommand{\gc}{\mathbf{g}}
\newcommand{\ar}{\overset{\circ}{\mathbf{a}}}
\newcommand{\acr}{\overset{\circ}{\mathrm{a}}}

\newcommand{\aci}{\bar{\mathrm{a}}}

\newcommand{\acc}{\mathrm{a}}

\newcommand{\br}{\overset{\circ}{{b}}}
\newcommand{\bcr}{\overset{\circ}{\mathrm{b}}}

\newcommand{\bi}{\bar{\mathbf{b}}}
\newcommand{\bci}{\bar{\mathrm{b}}}

\newcommand{\bcc}{\mathrm{b}}
\newcommand{\ccc}{\mathrm{c}}
\newcommand{\ccr}{\overset{\circ}{\mathrm{c}}}
\newcommand{\cci}{\bar{\mathrm{c}}}

\newcommand{\iphi}{\bar{\phi}}

\newcommand{\Gc}{{\bf G}}
\newcommand{\Gi}{\bar{\bf G}}
\newcommand{\Gcr}{\overset{\circ}{\mathrm{G}}}
\newcommand{\Gcc}{\mathrm{G}}
\newcommand{\Gci}{\bar{\mathrm{G}}}

\begin{document}
\title{Nonlinear morphoelastic energy based theory for stimuli responsive elastic shells}

\author{
\textsf{Matteo Taffetani$^{1}$ and Matteo Pezzulla$^{2}$}\\
\textit{$^{1}$ Institute for Infrastructure and Environment, School of Engineering,}\\
\textit{ The University of Edinburgh, Edinburgh EH9 3FG, United Kingdom}\\
\textit{$^{2}$Department of Mechanical and Production Engineering, }\\
\textit{{\AA}rhus University, {\AA}rhus N 8200, Denmark}
}

\date{}
\maketitle
\hrule\vskip 6pt
\begin{abstract}
Large deformations play a central role in the shape transformations of slender active and biological structures. A classical example is the eversion of the Volvox embryo, which demonstrates the need for shell theories that can describe large strains, rotations, and the presence of incompatible stimuli. In this work, a reduced two-dimensional morphoelastic energy is derived from a fully nonlinear three-dimensional formulation. The resulting model describes the mechanics of naturally curved shells subjected to non-elastic stimuli acting through the thickness, thereby extending previous morphoelastic theories developed for flat plates to curved geometries. Two representative constitutive laws, corresponding to incompressible Neo-Hookean and compressible Ciarlet–Geymonat materials, are examined to highlight the influence of both geometric and constitutive nonlinearities. The theory is applied to the eversion of open and closed spherical shells and to vesiculation processes in biological systems. The results clarify how compressibility, curvature, and through-the-thickness kinematics govern snap-through and global deformation, extending classical morphoelastic shell models. The framework provides a consistent basis for analyzing large deformations in elastic and biological shells driven by non-mechanical stimuli.
\end{abstract}
\vskip 6pt
\hrule


\vspace{1cm}
\section{Introduction}\label{SEC:Introduction}

Extreme deformations are a defining feature of morphing in active and biological slender structures, enabling them to perform their functions. These deformations typically involve both large strains and rotations. In active systems, shape changes are driven by the coupling between elastic responses and non-mechanical stimuli. A notable example is the eversion of the Volvox embryo \cite{Hohn2015}, which combines large rotations with active contraction, making it a reference problem for understanding the mechanics of morphing thin structures.

The mechanics of thin shells has been studied extensively, with an important distinction between fluid-like shells and solid-like shells. The former are characterized by the absence of energetic cost associated with the in-plane rearrangement of the material points; thus, their energy does not depend on in-plane shear deformations, although this does not imply that shear stresses cannot be sustained \cite{Dharmavaram2025,Sauer2017}. On the contrary, the energy associated with the latter depends on the full in-plane metric and not only on its determinant.

This work focuses on solid-like shells. A widely used constitutive model for passive slender shells is Koiter’s shell theory \cite{Koiter1961, Koiter1966}, which assumes small strains while retaining geometric nonlinearities associated with moderate rotations. Despite not being rigorously derived as a limit of three-dimensional elasticity (a step later achieved by Steigmann \cite{Steigmann2013}), its popularity stems from the resulting separation between area change and curvature change of the mid-surface in the energy functional proposed. As discussed by Ciarlet \cite{Ciarlet2005}, Koiter's model rests on two main assumptions: the Kirchhoff–Love kinematic hypothesis, geometrical in nature, and the plane-stress hypothesis, mechanical in nature. Later works have provided stronger theoretical foundations for these assumptions \cite{John1965, John1971}.

Deseri et al. \cite{Desseri2008} rigorously derived a reduced free energy establishing a consistent framework that connects local and non-local contributions and provides a solid basis for modeling biological and elastic membranes. For quadratic energy densities, Wood and Hanna \cite{Wood2019} showed that different strain measures yield different results. More recently, Vitral and Hanna \cite{Vitral2023,Vitral2023b} emphasized the presence of the intrinsic coupling between stretching, bending, and geometry, absent in Koiter’s model but guaranteed when reduction is performed from a fully nonlinear setting \cite{Steigmann2013}. To address the limitations of the classical assumptions, generalized Kirchhoff–Love hypotheses have been proposed \cite{Vitral2023,Ozenda2021,Stumpf1986}, allowing more general through-the-thickness responses. 

Morphing induced by activity can be incorporated in various ways. When the relaxation of elastic incompatibility is fast compared to the timescale of the stimulus, morphoelasticity \cite{Rodriguez1994,Goriely2017} provides an effective modeling framework, with stimuli introduced through energy functionals \cite{Kumar2023}. Its application to shells began with Efrati et al. \cite{Efrati2009}, who extended the linearized theory to non-Euclidean shells. Within this framework, Pezzulla et al. \cite{Pezzulla2018PRL} showed that non-elastic stimuli can act as boundary terms in the variational formulation. Later extensions included large rotations \cite{Haas2021} and fully nonlinear morphoelastic theories derived directly from three dimensions \cite{Sadik2016,Chen2025,Yu2025,Andrini2025}.

In this work, starting from a three-dimensional nonlinear morphoelastic theory, we derive a reduced two-dimensional energy functional that governs the morphing of slender, naturally curved objects. We focus on two classical constitutive laws: the incompressible Neo-Hookean material and the compressible Ciarlet–Geymonat model. By retaining geometric and constitutive nonlinearities without additional kinematic or constitutive assumptions, we include the effect of incompatible through-the-thickness stimuli, extending the work of Ozenda and Virga \cite{Ozenda2021} to curved morphoelastic shells. Our formulation is expressed in the language of differential geometry, with the non-elastic stimulus introduced in a general, phenomenological manner.

The paper is organized as follows. Section~\ref{Sec:NL_theory} presents the three-dimensional morphoelastic framework, based on the multiplicative decomposition of the deformation gradient. Section~\ref{Sec:Kinematics2D} introduces the shell kinematics and the dimensional reduction procedure. Section~\ref{Sec:ReducedEnergies} derives the reduced energy functionals for the chosen constitutive models. In Section~\ref{Sec:Numerics}, we apply the theory to spherical shells subject to curvature-inducing stimuli, highlighting the role of both geometrical and constitutive nonlinearities.

\section{Non-linear morphoelastic theory}\label{Sec:NL_theory}

\section*{Preliminaries}
We consider a three-dimensional body $\mathcal{B}_0$ embedded in the Euclidean space $\mathcal{E}$, described in terms of a curvilinear material coordinate system $[X^1, X^2, X^3]$. The position of the body in the reference configuration is given by a mapping $\mathbf{R}(X^i): \mathcal{B}_0 \rightarrow \mathcal{E}$ with $i \in \{1,2,3\}$. The associated reference metric is denoted by $\Gcr_{ij}$, with components defined as $\Gcr_{ij} = \gr_i \cdot \gr_j$, where the covariant tangent vectors are $\gr_i = \partial \mathbf{R} / \partial X^i$. Here and in the following, symbols in bold denote tensorial quantities while symbols in italics denote the component indicated by the subscripts.

The motion of the body is described by a mapping $\chi: \mathcal{B}_0 \rightarrow \mathcal{E}$, which determines the current configuration $\mathcal{B}$ so that the deformed position vector is given by $\mathbf{r} = \chi(\mathbf{R}((X^i)))$. The metric associated with the current configuration is denoted by $\Gcc_{ij}$, with components $\Gcc_{ij} = \gc_i \cdot \gc_j$, where $\gc_i = \partial \mathbf{r} / \partial X^i$ are the covariant tangent vectors in the deformed configuration.
The deformation gradient associated with the motion $\chi$ is defined as $\mathbf{F} = \nabla \chi$, where $\nabla$ denotes the three dimensional gradient with respect to the material coordinates $X^i$.

\begin{figure}
    \centering
    \includegraphics[width=0.9\textwidth]{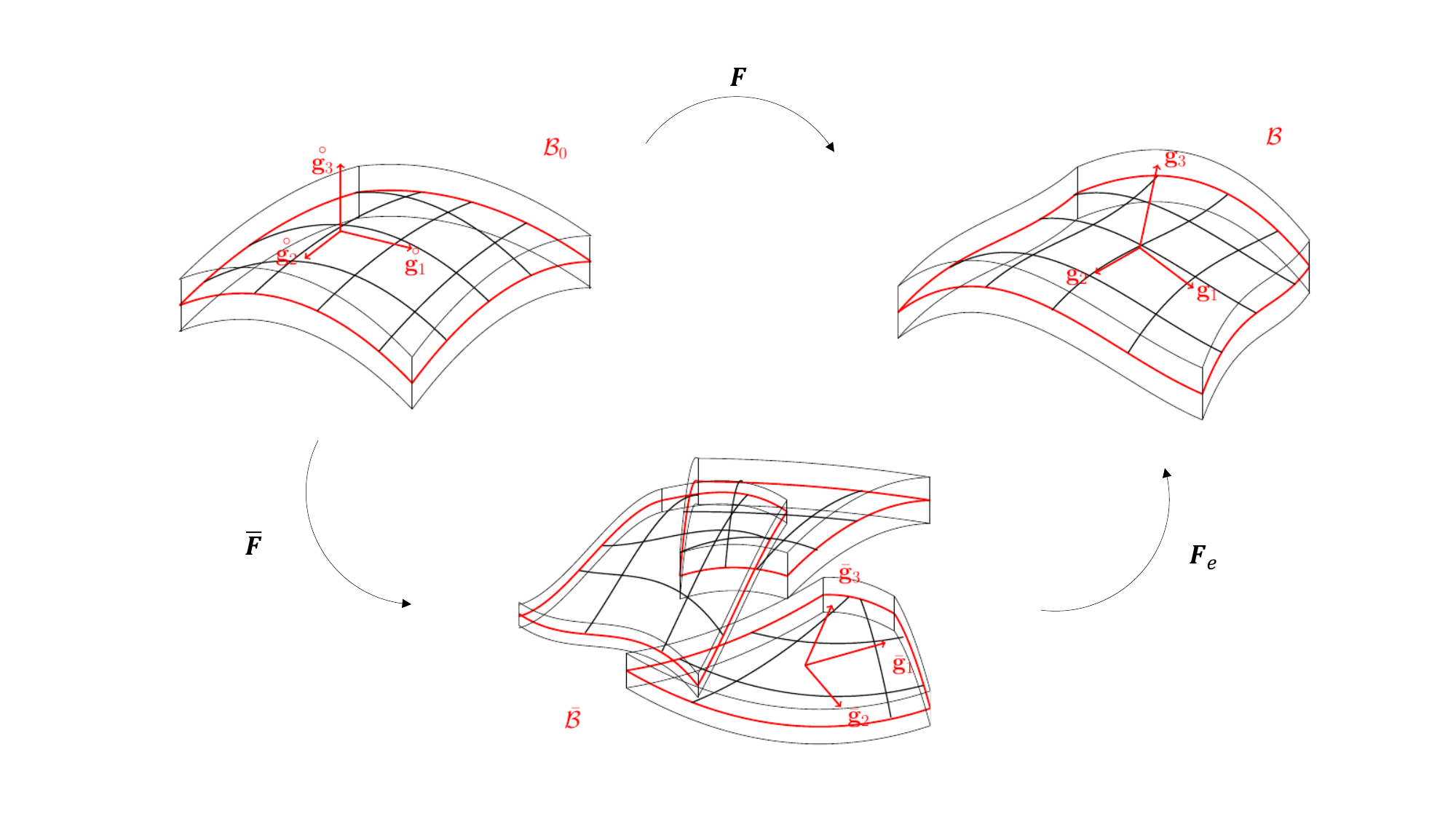}
    \caption{Multiplicative decomposition of the deformation gradient tensor. The total deformation gradient tensor $\mathbf{F}$ maps the reference configuration $\mathcal{B}_0$ to the current configuration $\mathcal{B}$. The presence of a non-mechanical stimulus is modelled via a deformation gradient tensor $\bar{\mathbf{F}}$ that maps the reference configuration $\mathcal{B}_0$ to the intermediate configuration $\bar{\mathcal{B}}$ that is not necessarily compatible, but where it is still possible to identify the tangent vectors associated with each point. The compatibility is guaranteed by the application of the map given by $\mathbf{F}_e$ from the intermediate to the current configuration from which it is possible to write the energy.}
    \label{FIG:ThreeConfigurations}
\end{figure}

Following the framework of morphoelasticity~\cite{Rodriguez1994, Goriely2017}, we introduce an intermediate configuration $\bar{\mathcal{B}}$, which may be geometrically incompatible, that is characterized by a prescribed metric $\Gci_{ij}$. This configuration accounts for the presence of a non-elastic stimulus applied to the system. As standard within this framework, we assume a multiplicative decomposition of the deformation gradient tensor as described in fig.~\ref{FIG:ThreeConfigurations} and given by 
\begin{equation}
\mathbf{F} = \mathbf{F}_e \, \bar{\mathbf{F}}\,,
\end{equation}
 where $\bar{\mathbf{F}}$ corresponds to the deformation associated with the incompatible metric $\Gci_{ij}$, and $\mathbf{F}_e$ describes the elastic part of the deformation. The elastic response of the material is assumed to depend solely on $\mathbf{F}_e$.

Before proceeding with the analysis of the energy, we introduce geometric and kinematic quantities that are relevant to develop our formulation. 
The mixed identity tensor ${\bf I}$ can be defined interchangeably as
\begin{equation}
    {\bf I} = \gi_i \otimes \gi^i = \gc_i \otimes \gc^i = \gr_i \otimes \gr^i,
\end{equation}
where repeating indices implies summation over $i$, and $\gi^i$, $\gc^i$, and $\gr^i$ denote the contravariant basis vectors of the intermediate, current, and reference configurations, respectively. We point out that, although the intermediate configuration is not associated with an embedding, it nonetheless describes a locally differentiable manifold so it is still possible to define tangent vectors. The deformation gradient tensors are defined as
\begin{equation}
    {\bf F} = \gc_i \otimes \gr^i, \quad {\bf F}_e = \gc_i \otimes \gi^i, \quad \bar{{\bf F}} = \gi_i \otimes \gr^i.
\end{equation}

Following the morphoelastic hypothesis adopted in this work, the elastic energy depends exclusively on the elastic deformation gradient ${\bf F}_e$. Thus we introduce the (elastic) right Cauchy–Green tensor
\begin{align}
    {\bf C}_e &= {\bf F}_e^T{\bf F}_e = \left(\gc_i \otimes \gi^i\right)^T \left(\gc_j \otimes \gi^j\right) = \left(\gi^i \otimes \gc_i\right)\left(\gc_j \otimes \gi^j\right) \notag \\
    &= (\gc_i \cdot \gc_j) \left(\gi^i \otimes \gi^j\right) = \Gcc_{ij} \left(\gi^i \otimes \gi^j\right),
\end{align}
where we emphasize that ${\bf C}_e$ is a tensor with covariant components living in the intermediate configuration.

The elastic energy $\Psi$ of an isotropic, hyperelastic material is, in general, written as a function of the three invariants of ${\bf C}_e$, i.e., $\Psi = \Psi(I_1, I_2, J_e)$, where $I_1 = \mathrm{tr}({\bf C}_e)$ is the first invariant, $I_2= (1/2) \left(\mathrm{tr}^2({\bf C}_e)-\mathrm{tr}({\bf C}^2_e)\right)$ is the second invariant, and $I_3 =\det({\bf C}_e)$ is the third invariant. It is useful to also define the Jacobian $J_e = \det({\bf F}_e) = \sqrt{I_3}$ as used in the analysis. Later in this work, we use energy formulations that are independent of the second invariant $I_2$, focusing on two classical models: the incompressible Neo-Hookean material, with energy density
\begin{align} \label{eq:NeoHookean3D}
    \Phi_{nh} = \frac{\mu^*}{2} (I_1 - 3) + p (J_e - 1)\,,
\end{align}
where $\mu^*$ is the shear modulus and $p$ is a Lagrange multiplier enforcing the incompressibility constraint, and the compressible Ciarlet–Geymonat material, with energy density \cite{Ozenda2021}
\begin{align} \label{eq:CiarletGeymonat3D}
    \Phi_{cg} = \frac{\mu^*}{2} \left(I_1 - 3 - \log I_3\right) + \frac{\lambda^*}{4} \left(I_3 - 1 - \log I_3\right)\,,
\end{align}
where $\lambda^*$ denotes the bulk modulus. We assume that the non-elastic stimulus does not alter the intrinsic material properties (e.g., $\mu^*$ and $\lambda^*$ remain constant), being consistent with a density-preserving non-elastic stimulus \cite{Lee2023}.

We want to express the invariants in terms of the current and intermediate metrics. The first invariant is written as
\begin{align}
    I_1 &= {\bf I}:{\bf C}_e = \gi_l \otimes \gi^l : \Gcc_{ij} \left(\gi^i \otimes \gi^j\right) = \Gcc_{ij} (\gi_l \cdot \gi^i)(\gi^l \cdot \gi^j) = \Gcc_{ij} \delta^i_l \Gci^{lj} = \Gcc_{ij} \Gci^{ij}\,,
\end{align}
where $\delta^i_l$ is the Kronecker delta and $\Gci^{ij}$ denotes the components of the inverse (contravariant) of the intermediate metric $\Gi_{ij}$. The intermediate metric describes a locally stress-free geometry. In relating the intermediate metrics and the tangent vectors in the intermediate configuration we use $\Gci_{ij}=\gi_i \cdot \gi_j$. The second invariant is written as 
\begin{align}
    I_2 &= \frac{1}{2}\left(\left({\bf I}:{\bf C}_e\right)^2 - {\bf I}:{\bf C}_e^2\right) = \frac{1}{2} \Bigl(\left(\Gcc_{ij} \Gci^{ij})^2\right) - \gi_l \otimes \gi^l : \left(\Gcc_{ik} \left(\gi^i \otimes \gi^k\right)\Gcc_{lj} \left(\gi^l \otimes \gi^j\right)\right)\Bigr)  \notag  \\
    & = \frac{1}{2} \left(\left(\Gcc_{ij} \Gci^{ij})^2\right) - \left(\Gcc_{ik}\Gcc_{lj} \Gci^{kl}\Gci^{ij}\right)\right)\,.
\end{align}
The third invariant, which captures the local volume change between the current and intermediate configurations, is given by
\begin{equation} \label{EQ:I3}
    I_3 = \det({\bf C}_e) = \frac{\det(\Gcc_{ij})}{\det(\Gci_{ij})}\,.
\end{equation}

Consequently, the volume change induced by the non-elastic stimulus alone, that is, from the reference to the intermediate configuration, is defined as
\begin{equation}
    J_m = \sqrt{\frac{\det(\Gci_{ij})}{\det(\Gcr_{ij})}}\,.
\end{equation}

Finally, the total elastic energy functionals corresponding to the two constitutive models introduced in equations \eqref{eq:NeoHookean3D} and \eqref{eq:CiarletGeymonat3D} are written as
\begin{equation}
\begin{aligned}
    \Psi_{nh} = \int_{\mathcal{B}_0} \Phi_{nh} \, J_m \sqrt{\det(\Gcr_{ij})}\, \mathrm{d}X^1 \mathrm{d}X^2 \mathrm{d}X^3, 
    \\
    \Psi_{cg} = \int_{\mathcal{B}_0} \Phi_{cg} \, J_m \sqrt{\det(\Gcr_{ij})}\, \mathrm{d}X^1 \mathrm{d}X^2 \mathrm{d}X^3\,.
    \end{aligned}
\end{equation}

For simplicity, we assume that our system is not subjected to body forces or surface tractions, although these contributions can be introduced in the model.

\section{3D to 2D reduction}\label{Sec:Kinematics2D}
Following the consistent finite-strain shell theory derived from 3D nonlinear elasticity as presented by, among others, Ciarlet~\cite{Ciarlet2005} and Steigmann~\cite{Steigmann2012, Steigmann2013}, we first present the kinematics of the shell and describe the geometry of the reference and current configurations.

The focus of our work is on slender curved objects, so we rewrite the reference domain $\mathcal{B}_0$ as $\mathcal{S}_0 \times [-h_0/2,h_0/2]$, where $\mathcal{S}_0$ defines the 2D midsurface of the object, and $h_0$ is its thickness. The slenderness hypothesis implies that, if $l_0$ indicates the characteristic in-plane length, then $h_0\ll l_0$.

It is convenient to introduce a set of curvilinear coordinates $[\eta^{1}, \eta^2, Z]$, where $\eta^\alpha$ (with $\alpha=1,2$) identify the in-plane coordinates, while $Z$ is the coordinate along the normal direction. Any point on the reference body can be identified with a position vector of the form
\begin{align}\label{eq:PositionReference}
    {\bf R}(\eta^{\alpha}, Z) = {\bf S}(\eta^{\alpha}) + Z {\bf N} (\eta^{\alpha})\,,
\end{align}
where ${\bf S}$ indicates the position on the midsurface and ${\bf N}$ indicates the unit normal to ${\bf S}$. This normal vector is, in general, a function of the position on the midsurface to account for an inhomogeneous non-zero natural curvature of the undeformed midsurface. 

The formalism employed in our formulation follows the works of Vitral and Hanna \cite{Vitral2023} and Ozenda and Virga \cite{Ozenda2021}. We carefully discuss where our approach diverges from theirs. In particular, we use the description proposed by Ozenda and Virga \cite{Ozenda2021} to relax the plane-strain hypothesis, allowing for changes of the thickness via the introduction of a function $\phi(\eta^{\alpha}, Z)$ in the definition of the current configuration as 
\begin{align}\label{eq:PositionCurrent}
    {\bf r}(\eta^{\alpha},Z) = {\bf s}(\eta^{\alpha}) + \phi(\eta^{\alpha}, Z) {\bf n} (\eta^{\alpha})\,,
\end{align}
where ${\bf s}$ is the position of a point in the current midsurface and ${\bf n}$ is the unit normal to the current midsurface ${\bf s}$. We assume that $\phi \sim O(h_0)$, so that it can be later expanded as a polynomial function in $Z$, with coefficients determined by minimizing a suitable functional, from either kinematical or energetic arguments. Compared to the work by Ozenda and Virga \cite{Ozenda2021}, our problem involves naturally curved shells also subjected to a non-elastic stimulus. The use of the kinematic assumption of Kirchhoff-Love type is justified by our interest in considering the mechanics of thin isotropic shells in bending- and membrane-dominated regimes, for which this kinematic relation can be formally derived via asymptotic expansion \cite{Ciarlet2000, Ciarlet2005}.

By direct computation, using the ansatz introduced in eq.~ \eqref{eq:PositionReference}, the 3D metric describing the body in the reference configuration takes the simplified form
\begin{equation}
    \Gcr_{ij} \upd X^i \upd X^j = \Gcr_{\alpha \beta} \upd \eta^{\alpha} \upd \eta^{\beta}+Z^2 \upd Z \upd Z\,,
\end{equation}
where the metric is decoupled in the in-plane contribution, denoted by $\Gcr_{\alpha \beta}$, with the Greek indices $\alpha,\beta = 1,2$, and the through-the-thickness contribution. The components of the metric associated with those two contributions take the form \cite{Ciarlet2000}
\begin{align}
    \Gcr_{\alpha \beta} = \acr_{\alpha \beta}(\eta^{\alpha}) - 2 Z \bcr_{\alpha \beta} (\eta^{\alpha}) + Z^2 \ccr_{\alpha \beta}(\eta^{\alpha})\,, \qquad \Gcr_{33} = 1\,,
\end{align}
where $\acr_{\alpha \beta}$, $\bcr_{\alpha \beta}$, and $\ccr_{\alpha \beta}$ are the components of the first, second, and third fundamental forms of the mid-surface $\mathcal{S}_0$, respectively. 

Similarly, by direct computation using the position field in eq.~ \eqref{eq:PositionCurrent}, the covariant metric for the current configuration can be written as
\begin{equation}\label{eq:3Dto2D_metric}
    \Gcc_{ij}\upd X^i \upd X^j =  \Gcc_{\alpha \beta} \upd \eta^{\alpha} \upd \eta^{\beta}+\phi^2 \upd Z \upd Z \,,
\end{equation}
with the in-plane and out-of-plane contribution given as
\begin{align}
    \label{eq:CovariantMetricCurrent}
    \Gcc_{\alpha \beta} &= \acc_{\alpha \beta}(\eta^{\alpha}) - 2\phi(\eta^{\alpha},Z) \bcc_{\alpha \beta}(\eta^{\alpha}) + \phi^2(\eta^{\alpha},Z) \ccc_{\alpha \beta}(\eta^{\alpha})\,, &\Gcc_{33} = \left(\phi'\right)^2\,,
\end{align}
where $\left(\cdot\right)'$ indicates differentiation with respect to $Z$ and $\acc_{\alpha \beta}$, $\bcc_{\alpha \beta}$, and $\ccc_{\alpha \beta}$ are the components of the first, second, and third fundamental forms of the mid-surface in the current configuration, respectively.
In deriving these expressions, we make use of two assumptions, highlighted in \cite{Ozenda2021} and justified by the slenderness of the shell, that allow us to simplify the analysis. Indeed, $\Gcc_{\alpha 3}, \Gcc_{3 \alpha} \ll \Gcc_{33}$ because $\phi(\eta^{\alpha},0)=0$, which implies $\nabla_{\eta} \phi(\eta^{\alpha},0)=0$, and the thickness is small enough to guarantee that $\nabla_{\eta} \phi(\eta^{\alpha},Z) \ll \partial \phi / \partial Z$. Furthermore, since we consider orientation-preserving mappings, we require $\phi(\eta^{\alpha},0)=0$ and $\partial \phi / \partial Z (\eta^{\alpha},0)>0$ that implies that the first coefficient in the (defined later) polynomial expansion of $\phi$ must be positive.

The contravariant form of the reference metric can be derived by inverting the covariant form and performing an asymptotic expansion in the coordinate $Z$ as
\begin{align}\label{eq:Go_contravariant}
    \notag \Gcr^{\alpha \beta} &= \acr^{\alpha \beta} + 2 Z \overset{\circ}{a}^{\alpha \gamma} \bcr_{\gamma \delta} \acr^{\delta \beta} + Z^2  \left(-\overset{\circ}{a}^{\alpha \gamma} \ccr_{\gamma \delta} \acr^{\delta \beta} + 4 \left(\acr^{\alpha \gamma} \bcr_{\gamma \delta}\right)^2\acr^{\delta \beta}\right) + O(Z^3)\,, \\
    \Gcr^{33} &= 1\,,
\end{align}
with $\acr^{\alpha \beta} = (\acr_{\alpha \beta})^{-1}$. The contravariant second and third fundamental forms are $\bcr^{\alpha \beta} = \acr^{\alpha \gamma} \bcr_{\gamma \delta} \acr^{\delta \beta}$ and $\ccr^{\alpha \beta} = \acr^{\alpha \gamma} \ccr_{\gamma \delta} \acr^{\delta \beta}$, so that the term proportional to $Z^2$ simplifies to $3 \ccr^{\alpha \beta}$.

Likewise, the contravariant form of the current configuration is derived by inverting the covariant form and using $\phi \sim O(h)$, thus obtaining an expansion of the form
\begin{align}
\Gcc^{\alpha \beta} = \acc^{\alpha \beta} + 2 \phi \acc^{\alpha \gamma} \bcc_{\gamma \delta} a^{\delta \beta} + 3 \phi^2 \acc^{\alpha \gamma} \ccc_{\gamma \delta} \acc^{\delta \beta} + O(\phi^3)\,, \qquad \Gcc^{33} = \left(\phi'\right)^{-2}\,,
\end{align}

\textbf{Remark.} The coefficient of $Z^2$ in eq.~ \eqref{eq:Go_contravariant} contains terms that scale with the square of the curvature. Consequently, for consistency, we must retain the contribution of the third fundamental form, which is of the same order as the square of the second fundamental form. In the linearized formulation, as the one based on the derivation proposed by Koiter \cite{Koiter1966}, the kinematic ansatz considered implies that the terms of order $O(Z^2)$ in the metric expansion contribute to the energy only at order $O(h^5)$. These contributions are therefore neglected in theories that retain only terms of order $O(h)$ and $O(h^3)$. In contrast, we will show that in our fully nonlinear setting, the order $O(Z^2)$ terms in the metric expansion contribute at order $O(h^3)$ to the energy, and thus cannot be ignored.\\

To model non-elastic stimuli in the mechanics of curved shells, we assume that the metric (not necessarily compatible) associated with the intermediate configuration has the same form as in eq.~ \eqref{eq:3Dto2D_metric} so that we write

\begin{equation}\label{eq:3Dto2D_metric_intermediate}
    \Gci_{IJ} \upd X^{I} \upd X^{J} = \Gci_{\alpha \beta} \upd \eta^{\alpha} \upd \eta^{\beta}+\iphi^2 \upd Z \upd Z\,,
\end{equation}
where
\begin{align}
    \label{eq:CovariantMetricIntermediate}
    \Gci_{\alpha \beta} &= \aci_{\alpha \beta}(\eta^{\alpha}) - 2\iphi(\eta^{\alpha},Z)\bci_{\alpha \beta}(\eta^{\alpha}) + \iphi^2(\eta^{\alpha},Z) \bar{c}_{\alpha \beta}(\eta^{\alpha})\,, &\Gci_{33} = \left(\iphi'\right)^2\,.
\end{align}
In the intermediate configuration, the components of the target forms $\aci_{\alpha \beta}$, $\bci_{\alpha \beta}$, and $\bar{c}_{\alpha \beta}$ describe the effect of the in-plane non-elastic stimulus, while the function $\iphi(\eta^{\alpha}, Z)$ is used to describe its effect across the thickness. It is important to note that the target forms do not have the same meaning as the fundamental forms introduced when we presented the expansion of the metrics in the reference and current configurations: they rather must be seen as coefficients of an appropriate asymptotic expansion of the intermediate metric and, as a consequence, $\bar{c}_{\alpha \beta}$ is not necessarily derived from $\aci_{\alpha \beta}$ and $\bci_{\alpha \beta}$, contrarily to the third fundamental form being dependent on the first and the second ones in the reference and current configurations that are both associated with an embedding. Later, to simplify the analysis, we introduce constraints on $\bar{c}_{\alpha \beta}$ and we leave the analysis of the general form to future work. Furthermore, we assume that the function $\iphi$ has all the properties of the function $\phi$: (i) it is of order $O(h)$ and (ii), although the non-elastic stimulus can be of a  generic form, it is necessarily associated to an orientation preserving transformation and thus also the first coefficient in the expansion for $\iphi$ must be positive.

Exploiting the smallness of the function $\iphi$ with respect to the characteristic in-plane length $l_0$, the contravariant forms of the intermediate configuration is derived by inverting the covariant form and performing an asymptotic expansion to obtain
\begin{align}\label{eq:Gab_contravariant}
    \Gci^{\alpha \beta} = \aci^{\alpha \beta} + 2 \iphi \aci^{\alpha \gamma} \bci_{\gamma \delta} \aci^{\delta \beta} + 3\iphi^2 \aci^{\alpha \gamma} \bar{c}_{\gamma \delta} \aci^{\delta \beta}  + O(\iphi^3)\,, \qquad \Gci^{33} = \left(\iphi'\right)^{-2}\,.
\end{align}

The form presented in eq.~\eqref{eq:Gab_contravariant} introduces a constraint on $\bar{c}_{\alpha \beta}$ because it is derived by forcing $\bar{c}_{\alpha \beta} =\bci_{\alpha \gamma}\aci^{\gamma \delta}\bci_{\delta \beta}$ and, from now on we force this requirement. This does not affect the characteristic of the intermediate configuration of being globally incompatible because, setting $\aci_{\alpha \beta}$ and $\bci_{\alpha \beta}$ arbitrarily, we still guarantee to impose a target metric that is not necessarily associated with an embedding. 

\textbf{Remark.} In the approach used in Yu and Chen \cite{Yu2025} to derive a two-dimensional model for morphoelastic shells, or in Moulton et al. \cite{Moulton2020} to derive a one-dimensional model for morphoelastic rods, the authors include the non-elastic stimulus by prescribing the deformation gradient $\bar{\bf F}$. Then, by performing a suitable asymptotic expansion of the fields in the slenderness and with respect to the mid-surface or the centerline, the reduced energy can be derived by minimizing a problem on the cross-sections \cite{Lestringant2020}. Our approach introduces the non-elastic stimulus by assuming a particular form of the incompatible metrics as shown in eq.~ \eqref{eq:3Dto2D_metric_intermediate}. This approach translates into the possibility to independently prescribe three quantities: the first and second fundamental forms $\aci_{\alpha \beta}$ and $\bci_{\alpha \beta}$, associated with the in-plane contribution of the stimulus, and the function $\iphi$, associated with the through-the-thickness contribution of the stimulus. The non-elastic stimuli are often difficult to model \emph{formally} because of the complexity of the physics that regulates their origin. We thus decide to consider a phenomenological approach that, although relying on a specific ansatz for the incompatible metrics as in eq.~ \eqref{eq:CovariantMetricIntermediate}, gives the possibility to explore a large variety of non-elastic stimuli. The two formalisms are equivalent and they are connected via the relation
\begin{equation}
   \bar{\mathbf{C}}=\Gci_{ij}\left(\gr^{{i}}\otimes\gr^{j}\right)\,,
\end{equation}
where one can see that it is possible to either prescribe $\bar{\mathbf{C}}$ (or $\bar{\bf F}$) or the metric $\Gci_{ij}$ and derive the other. We illustrate this by looking at a simple non-mechanical stimulus of an isotropic three dimensional dilation of factor $\xi$; we can write
\begin{equation}
    \bar{\mathbf{C}}=\xi^2\Gcr_{ij}\left(\gr^{{i}}\otimes\gr^{j}\right)\,,
\end{equation}
which directly implies
\begin{equation}
    \Gci_{ij}=\xi^2\Gcr_{ij}\,.
\end{equation}
Given the form in eq.~ \eqref{eq:CovariantMetricIntermediate},  it can be directly derived that
\begin{align}
 &\aci_{\alpha \beta}(\eta^{\alpha}) - 2\iphi\bci_{\alpha \beta}(\eta^{\alpha}) +  \iphi^2 \cci_{\alpha \beta}(\eta^{\alpha}) = \xi^2 \left(\acr_{\alpha \beta}(\eta^{\alpha}) - 2 Z \bcr_{\alpha \beta}(\eta^{\alpha}) +  Z^2 \ccr_{\alpha \beta}(\eta^{\alpha})\right) \\  
 &\left(\frac{\partial \iphi}{\partial Z}\right)^2 = \xi^2\,,   
\end{align}
from which we can readily derive the quantities in the intermediate configuration as
\begin{equation}\label{eq:example}
    \aci_{\alpha \beta} = \xi^2\acr_{\alpha \beta}, \quad  \bci_{\alpha \beta} = \xi \bcr_{\alpha \beta}, \quad \iphi=\xi Z\,,
\end{equation}
(and naturally $\cci_{\alpha \beta} = \ccr_{\alpha \beta}$). We note that this stimulus, chosen here for its simplicity, does not describe an incompatible intermediate configuration but rather gives rise to a realizable manifold in the Euclidean space without the need of any compatibility to be restored by the elastic contribution, provided that boundary conditions allow it. Inspection of eq.~\eqref{eq:example} shows that an isotropic dilation in curved shells, $\overset{\circ}{\mathrm{b}}_{\alpha \beta} \neq 0$, is not just pure stretching because you must include a constant through-the-thickness stretch that causes the curvature to also change \cite{Vitral2023}.

\subsection{Expansion of the invariants of $\bf{C}_e$}

In Section \ref{Sec:NL_theory} we presented the invariants expressed in terms of the metrics for the three configurations. Here we construct their asymptotic expansion as $h_0\rightarrow0$, assuming $\phi,\iphi \rightarrow 0$ in this limit.\\

\emph{The first invariant, $I_1 = {\bf I}:{\bf C}_e$ is expressed as}
\begin{align}\label{eq:FirstInvariant}
    \notag I_1 &= \left(\acc_{\alpha \beta}\aci^{\alpha \beta} + ( \phi')^2 (\iphi')^{-2} \right) - 2 \left(\phi \bcc_{\alpha \beta}\aci^{\alpha \beta}  - \iphi \acc_{\alpha \beta} \bci^{\alpha \beta}  \right)  \\
    &- \left(4 \phi \iphi \bcc_{\alpha \beta} \bci^{\alpha \beta} -3\iphi^2 \acc_{\alpha \beta} \cci^{\alpha \beta} -\phi^2 c_{\alpha \beta} \bar{a}^{\alpha \beta} \right) + O\left((\phi+\bar{\phi})^3\right)\,.
\end{align}

The expression for the first invariant is, in general, an infinite series that we truncate by neglecting terms of order $O\left((\phi+\bar{\phi})^3\right)$. If the non-elastic stimulus is such that $\bi_{\alpha \beta} = \bf{0}$, then the expression for $I_1$ is exact at the order we truncate here, as expected.

To derive this form, we first rewrite the first invariant in terms of the in-plane and out-of-plane contributions of the metrics as
\begin{align}
    I_1 &= \Gcc_{\alpha \beta} \Gci^{\alpha \beta} + \Gcc_{33} \Gci^{33}\,.
\end{align}
Then, recalling the covariant and contravariant forms of the metrics in eqs.~\eqref{eq:CovariantMetricCurrent} and \eqref{eq:CovariantMetricIntermediate}, we can write, truncated 
\begin{align}
    I_1 &= \Gcc_{\alpha \beta} \Gci^{\alpha \beta} + \Gcc_{33} \Gci^{33} \\
    &\notag= \left(\acc_{\alpha \beta} - 2 \phi \bcc_{\alpha \beta} + \phi^2 \ccc_{\alpha \beta}\right) \left( \aci^{\alpha \beta} + 2 \iphi \bci^{\alpha \beta} +3\iphi^2 \cci^{\alpha \beta} + O(\iphi^3)\right) + (\phi')^2 (\iphi')^{-2}\,,
\end{align}
and, by expanding the product, we obtain the desired form.\\

\emph{The third invariant, $I_3 = \det {\bf C}_e$ is expressed as} 
\begin{align}\label{eq:ThirdInvariant}
    \notag I_3 =& \left(\frac{\phi'}{\iphi'}\right)^2\frac{\det(\acc_{\alpha\beta})}{\det(\aci_{\alpha\beta})}\left[1 -4 \left(\phi H - \iphi \bar{H}\right) \right.\\
    &\left.+2 \left(2 H^2+K\right)\phi^2 + 2 \left(6 \bar{H}^2-\bar{K}\right) \iphi^2 -16 H \bar{H} \phi \iphi\right] + O\left((\phi+ \phi)^3\right)\,.
\end{align}
Here we assume that $\iphi' \sim O(\iphi)$ and $\phi' \sim O(\phi)$.

To derive this form, we first write the third invariant in terms of the in-plane and out-of-plane contributions of the metrics as
\begin{align}
    I_3 = \frac{\det \left(\Gcc_{ij}\right)}{\det\left(\Gci_{ij}\right)} = \left(\frac{\phi'}{\iphi'}\right)^2\frac{\det \left(\Gcc_{\alpha \beta}\right)}{\det\left(\Gci_{\alpha \beta}\right)}\,.
\end{align}
Then we rewrite the in-plane contribution of the determinant in terms of the fundamental and target forms as (see appendix \ref{SM:Third invariant} for detailed algebra of the derivation)
\begin{align}
    \det{\Gci_{\alpha\beta}} = \det(\aci_{\alpha\beta})\left[1 -4 \iphi \bar{H}  +\iphi^2 \left(2\bar{K}+  4\bar{H}^2\right) \right] + O(\iphi^3)\,,
\end{align}
\begin{align}
    \det{\Gcc_{\alpha\beta}} =\det(\acc_{\alpha\beta})\left[1 -4 \phi H  +\phi^2 \left(2K+  4H^2\right) \right] + O(\phi^3)\,,
\end{align}
where we introduced the mean curvature $H$ and Gaussian curvature $K$ in current configuration and the target mean curvature $\bar{H}$ and target Gaussian curvature $\bar{K}$ in the intermediate configuration. The quantities $\bar{H}$ and $\bar{K}$ are computed as
\begin{equation}
    \bar{H}=\frac{1}{2}\mathrm{tr}\left(\bci_{\alpha \beta}\aci^{\alpha \beta}\right), \quad \bar{K}=\frac{\det(\bci_{\alpha \beta})}{\det(\aci_{\alpha \beta})}\,.
\end{equation}
Although these definitions resemble algebraically the ones used to compute mean and Gaussian curvatures, we must not intend $\bar{H}$ and $\bar{K}$ as curvatures associated with an embedding. Finally, by taking their ratio and expanding the expression in the functions $\phi$ and $\iphi$ we get the desired form. 
Naturally we obtain
\begin{align}\label{eq:J}
    \notag J =&\left(\frac{\phi'}{\iphi'}\right)\sqrt{\frac{\det\left(\acc_{\alpha\beta}\right)}{\det\left(\aci_{\alpha\beta}\right)}}\left[1 -2 \left(\iphi \bar{H}-\phi H\right) \right.\\
    &\left.+ \iphi^2\left(4\bar{H}^2-\bar{K}\right) -4 \bar{H} H\phi \iphi +\phi K\right] + O\left((\phi+ \phi)^3\right)\,,
\end{align}
where we recognize the quantity 
\begin{equation}\label{eq:Js}
    J_s=\sqrt{\frac{\det\left(\acc_{\alpha\beta}\right)}{\det\left(\aci_{\alpha\beta}\right)}}
\end{equation}
that describes the area change of the mid-surface.

From the general expression in eq.~\eqref{eq:J}, we can derive the Jacobian $J_m$ of the transformation from the reference to the intermediate configuration as 
\begin{align}\label{eq:ThirdInvariantJm}
    \notag J_m =& \iphi'\sqrt{\frac{\det\left(\aci_{\alpha\beta}\right)}{\det\left(\acr_{\alpha\beta}\right)}}\left[1 -2 \left(Z \overset{\circ}{H}-\iphi \bar{H}\right) \right.\\
    &\left.+ Z^2\left(4\overset{\circ}{H}^2-\overset{\circ}{K}\right) -4 \overset{\circ}{H} \bar{H}\iphi Z +\iphi\bar{K}\right] + O\left((\phi+ \phi)^3\right)\,,
\end{align}
that is later needed to rewrite the domain of integration for the energy.\\

\emph{The second invariant, $I_2= (1/2) \left(\mathrm{tr}^2({\bf C}_e)-\mathrm{tr}({\bf C}^2_e)\right)$ is expressed as}
\begin{align}
    \notag I_2 =&J_s^2 + \left(\frac{\phi'}{\iphi'}\right)^2\acc_{\alpha \beta}\aci^{\alpha \beta}\\
    &\notag -2 \left[\phi \left(\left(\frac{\phi'}{\iphi'}\right)^2\bcc_{\alpha \beta}\aci^{\alpha \beta}+2J_s^2H\right)-\iphi\left(\left(\frac{\phi'}{\iphi'}\right)^2\acc_{\alpha \beta} \bci^{\alpha \beta} +2J_s^2 \bar{H}\right)\right]\\
    &\notag +\left[\phi^2\left( \left(\frac{\phi'}{\iphi'}\right)^2 \ccc_{\alpha \beta} \aci^{\alpha \beta}  +2J_s^2\left(2 H^2+K\right)\right)+\iphi^2\left( 3\left(\frac{\phi'}{\iphi'}\right)^2\acc_{\alpha \beta} \cci^{\alpha \beta} +2J_s^2\left(6 \bar{H}^2-\bar{K}\right)\right)\right.\\
    &\left.-4\phi\iphi\left(\left(\frac{\phi'}{\iphi'}\right)^2\bcc_{\alpha \beta} \bci^{\alpha \beta} +4 J_s^2H \bar{H} \right)\right]\,.
\end{align}

To derive this form, we first recall that, in our setting, the second invariant can be rewritten as shown in appendix \ref{SM:Second invariant}
\begin{equation}
    I_2 = I_3 \left(\frac{\phi'}{\iphi'}\right)^{-2} +\left(\frac{\phi'}{\iphi'}\right)^2 \left(I_1-\left(\frac{\phi'}{\iphi'}\right)^2\right).
\end{equation}
Hence it is straightforward to obtain the desired form by using the expressions in eq.~\eqref{eq:Js} and eq.~\eqref{eq:FirstInvariant}.\\

Formally, the forms of the invariants derived here are valid if the functions $\phi$ and $\iphi$ are regular enough. If we look for the functions $\phi$ and $\iphi$ having a polynomial expansion in $Z$ as 
\begin{align}\label{eq:ExpansionPhi}
    \phi &= \rho_1(\eta^{\alpha}) Z + \rho_2(\eta^{\alpha}) Z^2 +\rho_3(\eta^{\alpha}) Z^3 + O(Z^4)\,,
\end{align}
\begin{align}\label{eq:ExpansioniPhi}
    \iphi &= \bar{\rho}_1(\eta^{\alpha}) Z + \bar{\rho}_2(\eta^{\alpha}) Z^2 +\bar{\rho}_3(\eta^{\alpha}) Z^3 + O(Z^4)\,,
\end{align}
where we must require not only $\rho_1>0$ but also $\bar{\rho}_1 >0$ for the orientation-preserving constraint, we are guaranteed that the regularity requirement is satisfied. 

The form of the function $\phi$ is motivated by the choice in Ozenda and Virga \cite{Ozenda2021} and the remarks therein presented are valid also in our framework, with the addition of considering the presence of a natural curvature and a non-elastic stimulus. The polynomial expansion for $\iphi$ is justified by similarity with the choice for $\phi$. 

We note that the coefficients $\bar{\rho}_1$, $\bar{\rho}_2$ and $\bar{\rho}_3$ represent the through-the-thickness non-elastic stimulus and thus they can be arbitrary; on the other end, the coefficients $\rho_1$, $\rho_2$ and $\rho_3$ must be derived from either kinematic or mechanical considerations and, in the latter case, they are inevitably specific to the particular constitutive relationship used. Setting $\rho_1 = 1$ and $\rho_2=\rho_3=0$ in eq.~\eqref{eq:ExpansionPhi} is equivalent to imposing a plane-strain deformation. Equivalently, setting $\bar{\rho}_1 = 1$ and $\bar{\rho}_2=\bar{\rho}_3=0$ in eq.~ \eqref{eq:ExpansioniPhi} implies that the non-elastic stimulus is of a plane-strain form.

\section{Constitutive Modeling and Reduced Energies}
\label{Sec:ReducedEnergies}
The general kinematic description presented in Section~\ref{Sec:Kinematics2D} must be complemented with constitutive relationships specific to the problem under investigation. In what follows, we consider two representative materials: an incompressible Neo-Hookean material and a compressible Ciarlet--Geymonat model. These choices not only reflect widely used forms in the literature, but also enable validation of our derivation by recovering the known reduced model for non-morphoelastic plates from Ozenda and Virga~\cite{Ozenda2021}. 

\subsection{Incompressible Neo-Hookean Energy}

For an incompressible material, the coefficients of the through-the-thickness function in eq.~\eqref{eq:ExpansioniPhi} can be derived from kinematic considerations \cite{Ozenda2021}. Incompressibility implies $J = 1$ and we first limit ourselves to the case where the mid-surface is inextensible, i.e. $J_s = 1$, and we derive an energy functional where this latter constraint is enforced variationally. Inserting eq.~\eqref{eq:ExpansionPhi} and eq.~\eqref{eq:ExpansioniPhi} into eq.~\eqref{eq:J} yields the following conditions for the coefficients in the through-the-thickness function $\phi$ as 
\begin{align}
    \label{eq:TTT_coefficients_Incompressible-I}\rho_1(\eta^\alpha) &= \bar{\rho}_1\,,\\
    \rho_2(\eta^\alpha) &= \left(\bar{H}-H\right)\bar{\rho}_1^2 + \bar{\rho}_2,\\
    \label{eq:TTT_coefficients_Incompressible-III}\rho_3(\eta^\alpha) &= \bar{\rho}_3 + 2 \bar{\rho}_1 \bar{\rho}_2 \left(\bar{H}-H\right) + \frac{\bar{\rho}_1^3}{3}\left(6H\left(H - \bar{H}\right) - (K - \bar{K})\right)\,.
\end{align}
Upon integration over the thickness, also expanding the pressure field as $p = p_0(\eta^{\alpha}) Z + p_1(\eta^{\alpha}) Z + O(Z^2)$ (adding the term proportional to $Z^2$ would only introduce a redundant constraint at the truncation order used here), the energy functional in eq.~\eqref{eq:NeoHookean3D} becomes:
\begin{align}\label{eq:Energy_NH_IsoAreal}
    \Psi_{nc,2d} = \int_{\mathcal{S}_0} \left[ h_0 w^\textup{NH}_s + \frac{h_0^3}{12}w^\textup{NH}_b \right] \sqrt{\det(\aci_{\alpha \beta})} \, \upd \eta^1 \upd \eta^2 + O(h^5)\,,
\end{align}
where the integral is evaluated with respect to the coordinates in the reference configuration. For compactness, we present the quantities in the reduced functional when setting $\bar{\rho}_2=\bar{\rho}_3=0$. We refer to the appendix \ref{SM:NH_fullform} for their form including all the terms of the through-the-thickness stimulus. The stretching and bending terms, terms proportional to $h_0$ and $h_0^3$ respectively, are given by
\begin{align}
    w^\textup{NH}_s = &\bar{\rho}_1\left( \frac{\mu^*}{2} \left(\acc_{\alpha \beta} \aci^{\alpha \beta} -2\right) + p_0(J_s - 1) \right),\\
    \notag w^\textup{NH}_b = & \frac{\mu^*}{2}\bar{\rho}_1^{3} \left( 2\frac{\bar{K}}{\bar{\rho}_1}  w^\textup{NH}_s+ 2\left(H-\bar{H}\right) \bcc_{\alpha \beta} \aci^{\alpha \beta} -4 \bar{H} \acc_{\alpha \beta} \bci^{\alpha \beta} +  p_0 f_{p0}+p_1 f_{p1}\right.\\
    &\left.+  \left( - 4 \bcc_{\alpha \beta} \bci^{\alpha \beta} + \ccc_{\alpha \beta} \aci^{\alpha \beta} +  3\acc_{\alpha \beta} \cci^{\alpha \beta}\right) + f_1\right)\,,
\end{align}
with
\begin{align}
    f_1 &= \left(4 (H-\bar{H})(4H+\bar{H})-2K+2\bar{K}\right) ,\\
    f_{p0} &=\frac{24}{\mu^*}   H J_s (H-\bar{H})\,\\
    f_{p1} &=\frac{4}{\mu^*\bar{\rho}_1} \left(-2 H J_s+ \bar{H} J_s+\bar{H}\right)\,.
\end{align}

It can be checked that the constraint of incompressibility implies that the thickness does not change, at leading order, except for the effect of the applied non-elastic stimulus via $\bar{\rho}_1$
\begin{equation}
    h \simeq \int_{-h_0/2}^{h_0/2} \frac{\partial \phi}{\partial Z} \, \upd Z = \bar{\rho_1} h_0 + \frac{\rho_3}{4} h_0^3\,.
\end{equation}

It is possible to relax the constraint $J_s=1$ and derive an energy functional where the full constraint $J=1$ is imposed kinematically \cite{Ozenda2021}. Following the same procedure employed in the derivation of the expressions in eqs.~\eqref{eq:TTT_coefficients_Incompressible-I} -- \eqref{eq:TTT_coefficients_Incompressible-III} , the coefficients in the through-the-thickness function $\phi$ modify as 
\begin{align}\label{eq:TTT_coefficients_Incompressible} 
    \rho_1(\eta^\alpha) &= \frac{\bar{\rho}_1}{J_s},\\
    \rho_2(\eta^\alpha) &= \frac{\bar{\rho}_1^2}{J_s^2} \left(\bar{H}-H\right) + \frac{\bar{\rho}_2}{J_s},\\
    \rho_3(\eta^\alpha) &=\frac{\bar{\rho}_3}{J_s} + 2 \frac{\bar{\rho}_1 \bar{\rho}_2}{J_s^2}\left(J_s\bar{H}-H\right) + \frac{\bar{\rho}_1^3}{3J_s^3}\left(6H\left(H - J_s\bar{H}\right) - (K - J_s^2\bar{K})\right)\,.
\end{align} 
Upon integration over the thickness the terms in the energy in eq.~\eqref{eq:Energy_NH_IsoAreal} in this case becomes
\begin{align}
    w^\textup{NH}_s = &\frac{ \bar{\rho}_1\mu^*}{2} \left(\acc_{\alpha \beta} \aci^{\alpha \beta} -3 + \frac{1}{J_s^2}\right),\\
    \notag w^\textup{NH}_b = &\frac{\mu^*\bar{\rho}_1^3}{2 J_s^4}\left(J_s^4 \bar{K} w_s^{NH} + 2J_s\left(H-\bar{H} J_s\right) \bcc_{\alpha \beta} \aci^{\alpha \beta} -4J_s^4 \bar{H} \acc_{\alpha \beta} \bci^{\alpha \beta}\right. \\
    &\left.+ J_s^2 \left( - 4 J_s\bcc_{\alpha \beta} \bci^{\alpha \beta} + \ccc_{\alpha \beta} \aci^{\alpha \beta} +  3 J_s^2\acc_{\alpha \beta} \cci^{\alpha \beta}\right) \right.\\
    &\left.+ 2 \left(8 H^2 - 6 H \bar{H} J_s - K + J_s^2 \left(-2 \bar{H}^2 + \bar{K}\right)\right)\right)\,,
\end{align}
valid when imposing $\bar{\rho}_2=\bar{\rho}_3=0$.
\subsection{Compressible Ciarlet--Geymonat Energy}\label{Sec:CG_2D}

It is convenient to present the compressible Ciarlet--Geymonat material by rescaling the energy by the shear modulus $\mu^*$ to obtain
\begin{align}\label{eq:CiarletGeymonat2D}
    \left(\mu^*\right)^{-1}\Psi_{cg} = \int_{\mathcal{S}_0} \left[ h_0 w^{CG}_s + \frac{h_0^3}{12} w^{CG}_b \right] \sqrt{\det(\aci_{\alpha \beta})} \,\upd \eta^1 \upd \eta^2+ O(h^5)\,,
\end{align}
and we introduce the rescaled bulk modulus as $\lambda = \lambda^*/\mu^*$. For compactness, here we assume again that $\bar{\rho}_2=\bar{\rho}_3=0$ and we refer to the appendix \ref{SM:CG_fullform} for the full form of the reduced energy. The stretching contribution reads
\begin{align}\label{eq:CG_Stretching}
    \notag w^\textup{CG}_s = &\bar{\rho}_1 \left[\frac{1}{2} \left(\acc_{\alpha \beta} \aci^{\alpha \beta} + \left(\frac{\rho_1}{\bar{\rho}_1}\right)^2 - 3 - \log\left(\left(\frac{\rho_1}{\bar{\rho}_1}\right)^2 J_s\right) \right) \right.\\
    &\left.+ \frac{\lambda}{4} \left(\left(\frac{\rho_1}{\bar{\rho}_1}\right)^2 J_s - 1 - \log\left(\left(\frac{\rho_1}{\bar{\rho}_1}\right)^2 J_s\right) \right) \right],
\end{align}
and minimizing it with respect to $\bar{\rho}_1$ yields
\begin{equation}\label{eq:a}
    \rho_1 = \bar{\rho}_1 \sqrt{\frac{\lambda + 2}{\lambda J_s + 2}}.
\end{equation}

The bending contribution is given by:
\begin{align}\label{eq:CG_Bending}
    w_b^\textup{CG} &= \bar{\rho}_1^3 \bar{K} w^\textup{CG}_s -\bar{\rho}_1 \rho_2 \bcc_{\alpha \beta} \aci^{\alpha \beta} - 2 \bar{\rho}_1^2 \rho_1 \acc_{\alpha \beta} \bci^{\alpha \beta} \\
    \notag &- 2 \rho_1 \bar{\rho}_1^2 \bcc_{\alpha \beta} \bar{b}^{\alpha \beta} + \frac{1}{2} \rho_1^2 \bar{\rho}_1 \ccc_{\alpha \beta} \aci^{\alpha \beta} - \frac{3}{2} \bar{\rho}_1^3 \acc_{\alpha \beta} \cci^{\alpha \beta} + g_4.
\end{align}

where 
\begin{align}
    \notag g_4=& 
    \bar{\rho}_1 \left(\frac{\rho_2^2 (\lambda+2)}{\rho_1^2}+\frac{1}{2} \rho_1^2 \left(2 H^2 (\lambda+2 )+2 \bar{H}^2 J_s \lambda-\lambda (J_s \bar{K}+K)-2 K\right)\right.\\
    \notag &\left.+ \rho_2 H (\lambda+2)\right)+\bar{\rho}_1^2\frac{2 \bar{H}(\lambda+2) \left(\rho_2-\rho_1^2 H\right)}{\rho_1}+\frac{1}{2} \bar{\rho}_1^3 \left(2 \bar{H}^2+\bar{K}\right) (\lambda+2)\\
    \notag &+\frac{\bar{\rho}_1^{-1}}{2} \left(J_s \lambda \left(\rho_1^4 \left(2 H^2+K\right)-10 \rho_1^2 \rho_2 H+2 \rho_2^2\right)+4 \rho_2^2 \right)\\
    &-2 \rho_1 \bar{H} \left(\rho_1^2 H J_s  \lambda-\rho_2 J_s  \lambda+2 \rho_2\right).
\end{align}

Minimization of \eqref{eq:CG_Bending} with respect to $\rho_2$ leads to:
\begin{align}\label{eq:b}
    \rho_2 = \frac{1}{2} \bar{\rho}_1^2 \left[\left(\frac{H(\lambda + 2)(2 J_s \lambda - 1)}{(J_s \lambda + 2)^2} - \frac{2 \bar{H} J_s \lambda \sqrt{\lambda + 2}}{(J_s \lambda + 2)^{3/2}} \right) + \frac{\bcc_{\alpha \beta} \aci^{\alpha \beta}}{4 J_s \lambda + 8}\right].
\end{align}

Finally, it can be shown that, as in the non-morphoelastic plate case~\cite{Ozenda2021}, the function $\rho_3$ does not contribute to the energy up to $O(h_0^5)$ and thus remains undetermined.

\subsection{Linearization and bending-stretching coupling}\label{SEC:linearization}
It is illustrative to understand the behavior of our model against a linear elastic shell model, based on the Koiter energy with distortions, as used in \cite{Pezzulla2018PRL}. The Koiter shell model relies on a series of assumptions. Beyond the fact that its validity is limited to small strains, (i) it requires the use, in this exact order \cite{Efrati2009}, of the plane stress hypothesis to rewrite the constitutive relationship and the plane strain hypothesis to constrain the through-the-thickness behavior (when it is derived from a 3D setting \cite{Steigmann2013}) and (ii) in the energy formulation, stretching and bending strains are decoupled, with the former entering only at order $O(h_0)$ and the latter only at order $O(h_0^3)$.

Here we show that linearization in the limit of small strain and moderate rotations of our reduced energy functional does not reduce to the morphoelastic Koiter energy functional. Although the same arguments would apply for the general through-the-thickness modeling, to keep the notation more compact we perform this linearization assuming plane strain and no stimulus imposed through-the-thickness, that implies $\rho_1=\bar{\rho}_1=1$, $\rho_2=\rho_3=\bar{\rho}_2=\bar{\rho}_3=0$, to consider the closest scenario to the classic Koiter theory. We thus look for the expansion of the fundamental forms as
\begin{equation}
    \acc_{\alpha \beta} = \aci_{\alpha \beta} + \varepsilon \tilde{\mathrm{a}}_{\alpha \beta} \,, \quad \bcc_{\alpha \beta} = \bci_{\alpha \beta} + \varepsilon \tilde{\mathrm{b}}_{\alpha \beta} \,,
\end{equation}
with $\varepsilon \ll 1$, and we aim to linearize the Ciarlet--Geymonat reduced energy in \eqref{eq:CiarletGeymonat2D}. If we denote the components of the stretching strains as $\epsilon_{\alpha \beta}^s = \aci^{\alpha \gamma} \tilde{\mathrm{a}}_{\gamma \beta}$ and the components of the bending strains as $\epsilon_{\alpha \beta}^b = \aci^{\alpha \gamma} \tilde{\mathrm{b}}_{\gamma \beta}$, the energy up to order $O(h_0^3)$ takes the form
\begin{equation}
    \left(\mu^*\right)^{-1}\Psi_{\textup{lin}} = \int_{\mathcal{S}_0} \left( \frac{h_0}{8} w_s^{\textup{lin}} + \frac{h_0^3}{2} w_b^{\textup{lin}} + h_0^3 w^{\textup{gnl}} \right) \sqrt{\det(\aci_{\alpha \beta})} \, \upd \eta_1 \, \upd \eta_2 \,,
\end{equation}
where 
\begin{align}
    w_s^{\textup{lin}} &= \lambda \, \left(\epsilon_{\alpha \alpha}^s\right)^2 +  \,2\epsilon_{\alpha \beta}^s\epsilon_{\alpha \beta}^s \,, \\
    w_b^{\textup{lin}} &= \lambda \, \left(\epsilon_{\alpha \alpha}^b\right)^2 + 2  \,\epsilon_{\alpha \beta}^b\epsilon_{\alpha \beta}^b \,, \\
    \nonumber w^{\textup{gnl}} &= \bar{H}  \left( \epsilon_{\alpha \alpha}^s \left(\epsilon_{\alpha \beta}^s\bar{S}^{\alpha}_{\beta}\right) - \epsilon_{\alpha \gamma}^s\epsilon_{\gamma \beta}^b\epsilon_{\alpha \gamma}^s\epsilon_{\gamma \beta}^b \right) \\
    \nonumber & \quad + \frac{1}{8} \bar{K} \left( -6  \,\epsilon_{\alpha \beta}^s\epsilon_{\alpha \beta}^s - 4  \, \epsilon_{\alpha \alpha}^s \left(\epsilon_{\alpha \alpha}^s - 3 \, \tilde{b}_{\alpha \gamma}\left(\bar{S}^{-1}\right)^{\alpha}_{\beta} \right)  + \lambda \, \epsilon_{\alpha \alpha}^s \left( -3 \epsilon_{\alpha \alpha}^s + 4 \, \tilde{b}_{\alpha \gamma}\left(\bar{S}^{-1}\right)^{\alpha}_{\beta} \right) \right) \\
    & \quad + \frac{1}{2} \left(\epsilon_{\alpha \beta}^s \bar{S}^{\alpha}_{\beta}\right)  \left( \left( \epsilon_{\alpha \beta}^s \bar{S}^{\alpha}_{\beta}  - 3 \, \epsilon_{\alpha \alpha}^b \right) + \lambda \left(\epsilon_{\alpha \beta}^s \bar{S}^{\alpha}_{\beta}  - 2 \, \epsilon_{\alpha \alpha}^b \right) \right) \,,
\end{align}
where we introduce the symbol $\bar{S}^{\alpha}_{\beta}= \aci^{\alpha \gamma} \bci_{\gamma \beta}$ that identifies a generalized target shape operator. 

This linearization reveals two aspects of the behavior of our model in the presence of small strains and moderate rotations. If the undeformed configuration is flat, bending and stretching are fully decoupled, although with different coefficients due to the lack of plane stress enforcement (for comparison, the reader can check the formulation used, for example, in Pezzulla et al. \cite{Pezzulla2018PRL}); if the undeformed configuration is curved, the term $w_{\textup{gnl}}$ is non-zero and the non-linearities due to the interaction with the geometry are retained \cite{Steigmann2013,Wood2019}. 

\section{Examples: Non-linear morphing of curved shells}\label{Sec:Numerics}
In our examples, we focus on axisymmetric morphologies and we consider the Ciarlet-Geymonat constitutive relationship to also investigate the effect of compressibility in the morphing strategies adopted. We consider a spherical cap of undeformed radius $R_0$ and thickness $h_0$ that deforms into shells of revolution of arbitrary profile. In a Cartesian reference system denoted by the unit vectors $[\mathbf{e}_x,\mathbf{e}_y,\mathbf{e}_z]$, the mid-surface in the initial configuration can be parametrized as
\begin{equation}
    {\bf R}(s, \omega) = R_0\sin(s) \cos(\omega) \mathbf{e}_x + R_0\sin(s) \sin(\omega) \mathbf{e}_y + R_0\cos(s) \mathbf{e}_z\,,
\end{equation}
where $s \in [-\Theta,\Theta]$ is the meridional coordinate and $\omega \in [0,\pi]$ is the azimuthal coordinate where $\Theta$ indicates the opening angle. Mean and Gaussian curvatures of the initial configuration are given by $ \overset{\circ}{H}= - R_0^{-1}$ and $ \overset{\circ}{K}=R_0^{-2}$.

\subsection{Minimization of the energy functional}
The energy functional derived in Section \ref{Sec:CG_2D} is implemented in COMSOL Multiphysics, following the same procedure presented in Pezzulla et al. \cite{Pezzulla2018PRL}. Both the stretching (the term proportional to $h_0$) and bending (the term proportional to $h_0^3$) components in the energy are functions of the first, second, and third fundamental forms. In turn, the fundamental forms can be expressed as functions of the embeddings of the undeformed and deformed mid-surfaces, namely $\mathbf{X}$ and $\mathbf{x}$. As a result, the total elastic energy of the shell becomes a second-order functional of $\mathbf{x}$, which we minimize in COMSOL Multiphysics.

\subsection{Eversion of a cap}
Pezzulla et al. \cite{Pezzulla2018PRL} investigated the eversion of a spherical cap, whose mid-surface in the deformed configuration is parametrized as
\begin{equation}
    {\bf r}(s, \omega) = \phi(s) \cos(\omega) \mathbf{e}_x + \phi(s) \sin(\omega) \mathbf{e}_y + \psi(s) \mathbf{e}_z\,,
\end{equation}
under the non-elastic stimulus defined by 
\begin{equation}\label{eq:stimulus}
    \aci_{\alpha \beta} = \acr_{\alpha \beta}, \quad \bci_{\alpha \beta} = \bcr_{\alpha \beta} + \kappa \acr_{\alpha \beta}, \quad \bar{\rho}_1=\rho_1, \quad \bar{\rho}_2=\bar{\rho}_3=0.
\end{equation}
This stimulus imposes a change the curvature of size $\kappa$, the only controlling parameter, without affecting the area of the mid-surface fixed, and without adding any contribution through-the-thickness. This stimulus also implies $\cci_{\alpha \beta} = 2 \bar{H} \bci_{\alpha \beta} - \bar{K}\aci_{\alpha \beta}$  where $\bar{K}=\overset{\circ}{K} + \kappa^2 - 2 \kappa R_0^{-1}$ and $\bar{H} = \overset{\circ}{H} + \kappa$, from straightforward algebra and consistently with the constraint on $\cci_{\alpha \beta}$ over which we derive the energy functionals. The energy is minimized by fixing to zero the displacement of the apex of the cap at $s=0$ and tracking the vertical component  $w$ of the displacement of the free end of the cap at $s=\Theta$ . 

We first investigate the effect of imposing the plane strain assumption within our framework, that translates into setting $\rho_1=1$ and $\rho_2=0$.

\begin{figure}
    \centering

        \begin{tikzpicture}

    	 \node[inner sep=0pt] at (1.2,6.67){
    \includegraphics[width=0.4\textwidth]{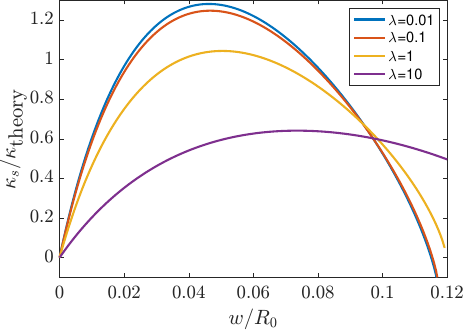}};

    	 \node[inner sep=0pt] at (8.4,6.67){
    \includegraphics[width=0.4\textwidth]{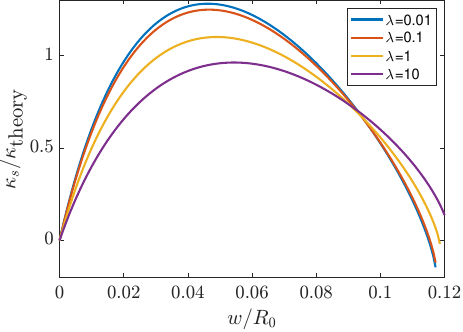}};

   \node[inner sep=0pt] at (-2.1,9.4){
    a)};
   \node[inner sep=0pt] at (5,9.4){
    b)};
    
    \end{tikzpicture}
    
    \caption{Snapping of a spherical cap. Natural curvature at snapping, normalized by the theoretical prediction from \cite{Pezzulla2018PRL}, as a function of the vertical displacement at the free end $w$, normalized by the radius of the shell $R_0$, for $\Theta=\pi/6$ and $\lambda=0.01,0.1,1,10$ when (a) $\rho_1=1$ and $\rho_2=0$ and (b) $\rho_1$ and $\rho_2$ fixed by eqs.~\eqref{eq:a} and \eqref{eq:b}.}
    \label{fig:cap_ab10}
\end{figure}
Fig.~\ref{fig:cap_ab10} shows the curvature stimulus at snap-through $\kappa_s$, normalized by the theoretical prediction from Pezzulla et al. \cite{Pezzulla2018PRL} obtained by using the classic morphoelastic Koiter shell theory \cite{Efrati2009}, function of the normalized displacement at the apex, $w/R_0$. In fig.~\ref{fig:cap_ab10}(a) we see that snap-through (indicated by the presence of a maximum in the curves) is present any value of $\lambda$ but it decreases when moving towards the limit of incompressibility, $\lambda \rightarrow\infty$, suggesting the possibility for snapping to be eventually suppressed. Experiments \cite{Pezzulla2018PRL} however, shows that snap-through is expected for morphing spherical caps also of incompressible (or nearly incompressible) materials. If we relax the plane strain requirement, thus allowing $\rho_1$ and $\rho_2$ to be fixed by eqs.~\eqref{eq:a} and \eqref{eq:b}, fig.~\ref{fig:cap_ab10}(b) shows how snap-through happens for any value of the shear to bulk moduli ratio and it remains close to the experimental values.

\begin{figure}
    \centering
    \begin{tikzpicture}

    \node[inner sep=0pt] at (8.3,6.7){
    \includegraphics[width=0.45\textwidth]{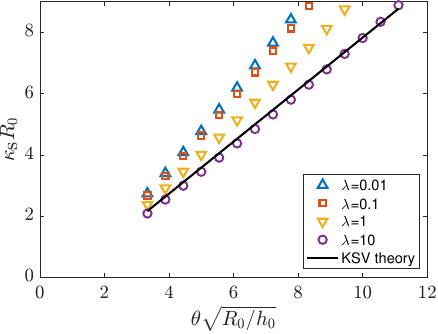}};

    \end{tikzpicture}
    \caption{Dimensionless imposed curvature at snapping, $\kappa_s R_0$ in terms of the rescaled shallowness $\theta\sqrt{R_0/h_0}$. The solid line represents the theoretical prediction from \cite{Pezzulla2018PRL}, valid for an incompressible Koiter shell model. The colored curves represent the result of our model for different values of $\lambda=0.01,0.1,1,10$ when $\rho_1$ and $\rho_2$ are fixed by eqs.~\eqref{eq:a} and \eqref{eq:b}.}
    \label{fig:cap_abgeneral}
\end{figure}

If we plot the dimensionless imposed curvature at snap-through, $\kappa_s R_0$, with the respect to the normalized opening angle as shown in fig.~\ref{fig:cap_abgeneral}, we see that the our model (i) recovers the absence of snapping for shallow shells, in rescaled form, $\Theta\left(h_0^{-1}R_0\right)^{1/2} \lessapprox 3.1$ and (ii) it is consistent with the prediction based on the morphoelastic Koiter theory for a incompressible material, that in turn has already been shown to correctly model the experiments. Furthermore, we show that snap-through requires a larger natural curvature in more compressible material because the compressibility partially accommodates the imposed geometrical frustration that is the root cause of a snapping event.

Although the morphoelastic Koiter theory has been proved to correctly model the snap-through events in spherical caps \cite{Pezzulla2018PRL}, the combination of an appropriate linearized theory that retains geometric non-linearities (in case of small strains and moderate rotations our model reduced to the form shown in Section \ref{SEC:linearization}) and the appropriate modeling of the through-the-thickness behavior may be the correct approach to be used to unveil the fundamental aspects in the physics of eversion morphing, also allowing the correct modelling of plane-strain and plane-stress scenarios.

\subsection{Eversion of a closed sphere}
Inspired again by the analysis in Pezzulla et al. \cite{Pezzulla2018PRL}, we look at the effect of the same incompatible non-elastic stimulus in eq.\eqref{eq:stimulus} on a closed sphere ($s \in [-\pi,\pi]$).

Defining $w_R$ as the homogeneous radial displacement, we impose the ansatz of the form
\begin{equation}
    {\bf \Phi}(s, \omega) = \left(R_0+w\right)\sin(s) \cos(\omega) \mathbf{e}_x + \left(R_0+w\right)\sin(s) \sin(\omega) \mathbf{e}_y + \left(R_0+w\right)\cos(s) \mathbf{e}_z\,,
\end{equation}
that describe a homogeneous expansion or contraction. We insert this ansatz in the energy in eq.~\eqref{eq:CiarletGeymonat2D} and we minimize it against the radial displacement $w_R$. Imposing the plane strain condition, $\rho_1=0$ and $\rho_2=0$ and in absence of a through-the-thickness non-elastic stimulus, by expanding for $\kappa h_0 \ll 1$ we find that the radial displacement varies quadratically with the imposed curvature as
\begin{equation}
    \frac{w_R}{R_0} =\frac{1-\lambda}{24\left(1+\lambda\right)} \left(\kappa h_0\right)^2 + O((\kappa h_0)^3).
\end{equation}

\begin{figure}
    \centering
    \includegraphics[width=0.5\textwidth]{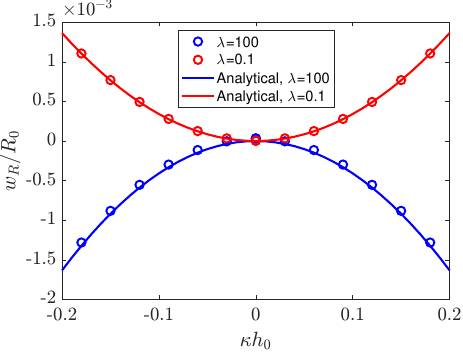}
    \caption{Homogeneous dimensionless displacement $w_R/R_0$ as a function of the dimensionless imposed curvature $\kappa R_0$, in the case of plane strain conditions. For $\lambda=100>1$ the shell deflates while for $\lambda=0.1<1$ the shell inflates.}
    \label{fig:enter-la}
\end{figure}

As shown in fig.~\ref{fig:enter-la} for two releveant cases, our model thus predicts that a homogeneous imposed curvature leads to a contraction of the sphere for any value of $\kappa$ imposed regardless of its sign, if the shell has $\lambda>1$, whereas extremely compressible materials with $\lambda<1$ would suffer an expansion for any value of $\kappa$ imposed; for the threshold value $\lambda=1$, any effect of the stimulus is exactly compensated by the compressibility, thus resulting in a shell that doesn't change radius. 
Interestingly, relaxing the plane strain hypothesis and thus retaining through-the-thickness behavior via the appropriate form of $\rho_1$ and $\rho_2$, leads to a radial displacement of the form
\begin{equation}
    \frac{w_R}{R_0} =\frac{3 \left(h_0/R_0\right) \lambda (\lambda+2)}{\left(h_0/R_0\right)^2 \left(10 \lambda^2+\lambda-2\right)+12 (\lambda+2)^2} (\kappa h_0)+ O\left((\kappa h_0)^2\right) 
\end{equation}
that allow us to recover (except for the sign) the linear behavior in $\kappa h_0$ highlighted in Pezzulla et al. \cite{Pezzulla2018PRL}

This analysis allows us to reiterate the importance of retaining the geometrical non-linearities when studying the limit of small strains. In our model, the issue of retaining a bending-to-stretching coupling, recently rediscussed by Wood and Hanna \cite{Wood2019}, can be seen in Section \ref{SEC:linearization} where we show that the coupling term identified by $w^{gnl}$ is identically zero only for intrinsically flat slender objects.

\subsection{Large deformation in budding and vesiculation}

Shapes associated with budding and vesiculation are typical during deformations in biological membranes and their morphologies affect their functionalities. Although the biochemical complexity behind budding in biological membranes is beyond the scope of this work, the mechanical model proposed here can be used to deepen the understanding of the role of large deformations and localized stimuli on the morphing strategies in active and biological solid-like curved shells as investigated in recent papers \cite{Chen2025, Yu2025}. Localized control of curvature and/or area growth are indeed strategies used by biological membranes in creating protrusions that can be precursor of vesiculation \cite{Willy2021,Zimmerberg2006}. Biological membranes are often modeled as fluid membranes \cite{Helfrich1973} because its microscopic structure of a lipidic bi-layer allows distances to be rearranged freely and thus it cannot sustain shear loads, differently from what happens in a solid elastic membrane. However, it has been shown that elastic features must be taken into account to fully characterize the experimental evidence on the mechanical behavior of biological membranes \cite{Deseri2013}; in those systems, the elastic behavior is thought to be due to the connection of the membranes to an underlying protein mesh \cite{Lachowski2022,Janmey2006}. 

Supported by experimental data, Willy et al. \cite{Willy2021} proposed that the formation of a vesicle from a shallow substrate is controlled by a combined in-plane growth followed by a localized curvature stimulus. Because this morphing strategy is within the class of stimuli most naturally accommodated within our modelling framework, we aim to qualitatively test what morphologies emerge in a non-linear morphoelastic spherical cap when we apply simple localized stimuli in curvature and metric, leaving a more detailed analysis to future studies. 

We consider an axisymmetric shallow spherical shell with radius $R_0=0.25$ and thickness $h_0=0.005$, so that thickness-to-radius ratio is $h_0/R_0=0.02$, and opening angle $\Theta=\pi/8$, subjected to a prescribed morphing non-elastic stimulus imposed around the apex in a domain $\mathcal{I}_s=\{(s,\omega): s \in [-\Theta/4,\Theta/4] , \omega  \in[0,\pi] \}$. The dimensionless bulk modulus is $\lambda = 7.5$. The stimulus is constant in the domain $\mathcal{I}_s$ (and identically zero outside $\mathcal{I}_s$) and takes the form (where we set 1 to indicate the meridional direction and 2 to indicate the azimuthal direction)
\begin{align}\label{eq:stimulusII}
    \notag \bar{a}_{11} = \left(1+ \frac{1}{5}\gamma_{a1}\bar{\kappa}\right)\ar_{11}, \quad \bar{a}_{22} = \left(1+  \frac{1}{5}\gamma_{a2}\bar{\kappa}\right)\ar_{22}, \\
    \bar{b}_{11} = \left(1+ \gamma_{b1}\bar{\kappa}\right)\br_{11}, \quad \bar{b}_{22} = \left(1+  \gamma_{b2}\bar{\kappa}\right)\br_{22}, 
\end{align}
where the quantities $\gamma_{a1}, \gamma_{as}, \gamma_{b1}, \gamma_{b2}$ are constant. We set the anisotropy ratios as $\eta_a=\gamma_{a1}/\gamma_{a2}$ and $\eta_b=\gamma_{b1}/\gamma_{b2}$ and we investigate their role in the emerging morphologies when increasing the value of the single controlling parameter $\bar{\kappa}$ from zero. To deal with a set of non-elastic stimuli of \emph{equivalent magnitude} we impose the same L2-norm for both the curvature and metric stimuli in all the cases addressed, while varying the anisotropy ratios: this means that we set $\eta_a=0.63/1.26=0.5$, $\eta_a=1/1=1$ and $\eta_a=1.26/0.63=2$ (and the same applies for $\eta_b$). We do not prescribe any through-the-thickness stimulus but we maintain the correct through-the-thickness behavior by including the form of $\rho_1$ and $\rho_2$ as in eq.~\eqref{eq:a} and eq.~\eqref{eq:b}. We set clamped boundary conditions at the outer edge of the spherical cap.

\begin{figure}
    \centering
    \includegraphics[width=0.9\textwidth]{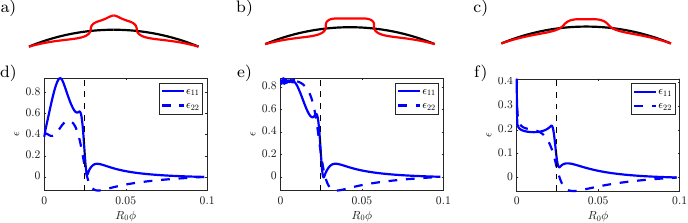}
    \caption{Morphologies obtained setting $\eta_a=0$ and (a) $\eta_b=0.5$, (b) $\eta_b=1$ and (c) $\eta_b=2$ for $\bar{\kappa}=300$ (panel (a) and (b)) and $\bar{\kappa}=128$ (panel (c)) . Meridional $\epsilon_{11}$ (continuous curves) and azimuthal (dashed curves) $\epsilon_{22}$ strain for a morphoelastic cap with $\eta_a=0$ and (d) $\eta_b=0.5$, (e) $\eta_b=1$ and (f) $\eta_b=2$ for $\bar{\kappa}=300$ (panel (d) and (e)) and $\bar{\kappa}=128$ (panel (f)).}
    \label{FIG:CurvatureOnly}
\end{figure}

We first explore the case in which only a curvature target is imposed, i.e. we set $\gamma_{a1}=\gamma_{a2}=0$. In fig.~\ref{FIG:CurvatureOnly}(a), \ref{FIG:CurvatureOnly}(b) and \ref{FIG:CurvatureOnly}(c), we show the emerging morphologies that are associated with $\eta_b=0.5$ and $\eta_b=1$ for $\bar{\kappa}=300$, and $\eta_b=2$ for $\bar{\kappa}=128$, respectively. A target curvature that imposes a larger stimulus in the meridional direction gives rise to an elongated bulge, similar to the morphologies observed at the onset of a protrusion. An isotropic stimulus generates a more symmetric bulge, although flattened around the apex. When the target field prescribes a larger curvature along the azimuthal direction, the morphology evolves toward a ring-like bulge (in fig.~\ref{FIG:CurvatureOnly}(c) we see the onset of a trough at the apex), with a morphology reminiscent  of the pits induced by clarthrin coats \cite{Willy2021} or a caveolar superstructure \cite{Golani2019}; imposing a larger curvature in the azimuthal direction has the effect of forcing the domain $\mathcal{I}_s$ to \emph{shrink} towards the apex more than the meridional contribution of the stimulus can accommodate, making the formation of a ring energetically favorable over a bulge. Interestingly, the emerging morphologies display regions of both positive and negative local curvature, although the target metrics contain only positive local curvatures. Examining the elastic strains associated with the three cases, as shown in  fig.~\ref{FIG:CurvatureOnly}(d), \ref{FIG:CurvatureOnly}(e) and \ref{FIG:CurvatureOnly}(f), we note the presence of both common and distinct features. Compressive azimuthal strains (dashed curves) emerge right outside $\mathcal{I}_s$ in all the cases, suggesting the potential for an instability in the hoop direction, possibly acting as a precursor to fracture \cite{Riccobelli2024}; the meridional strain (continuous curves) remains tensile throughout the domain $\mathcal{I}_s$ but exhibits a sharp transition at the boundary of the region where the stimulus is applied. Within the domain $\mathcal{I}_s$, the strains are always tensile; however, if their maximum is always at the apex for $\eta_b=1$ and $\eta_b=2$ (with a strong localization at the apex present when the ring-like morphology), when a larger curvature is imposed along the meridional direction ($\eta_b=0.5$), the maximum strain is found around the middle of the meridional extent of the stimulated domain $\mathcal{I}_s$, potentially being the cause for the emergence of a protruded shape, instead of the more flattened apex as seen when $\eta_b=1$.


\begin{figure}
    \centering
    \includegraphics[width=0.4\textwidth]{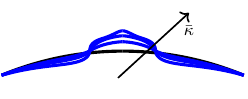}
    \caption{Evolution of shapes obtained for increasing values of $\bar{\kappa}=[64, 160,300]$ when the anisotropy ratios are set as $\eta_a=0$ and $\eta_b=0.5$.}
    \label{FIG:BulgeLikeA}
\end{figure}

It is interesting to see the evolution of the protrusion-like shape obtained setting $[\eta_a,\eta_b]=[0,0.5]$ and shown in fig.~\ref{FIG:BulgeLikeA} for increasing values of $\bar{\kappa}$. The protrusion, also associated with the presence of an inflection point where curvature changes sign, appears only at a later stage, whereas the initial stages of deformation accommodate more dome-like shapes.

The addition of a non-elastic stimulus acting on the metric, i.e. imposing a non-compatible area target, enriches the variety of the emerging shapes and their evolution for increasing values of $\bar{\kappa}$, even in the simple form of the non-elastic stimulus considered here. We limit the analysis to one representative case.

\begin{figure}
    \centering
    \includegraphics[width=0.9\textwidth]{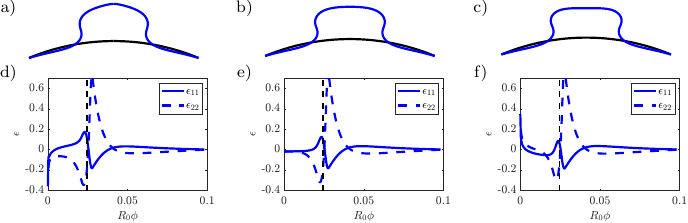}
    \caption{Morphologies obtained setting $\eta_b=1$ and (a) $\eta_a=0.5$, (b) $\eta_a=1$ and (c) $\eta_a=2$ for $\bar{\kappa}=200$ (panel (a) and (b)) and $\bar{\kappa}=76$ (panel (c)). Meridional $\epsilon_{11}$ (continuous curves) and azimuthal (dashed curves) $\epsilon_{22}$ strain for a morphoelastic cap with $\eta_b=1$ and (d) $\eta_a=0.5$, (e) $\eta_a=1$ and (f) $\eta_a=2$ for $\bar{\kappa}=200$ (panel (d) and (e)) and $\bar{\kappa}=76$ (panel (f)).}
    \label{FIG:RingLikeA}
\end{figure}

We showed that, when $\eta_a=0$ and $\eta_b=1$ a dome-like morphology emerges. Fig.~\ref{FIG:RingLikeA}, shows that this morphology is not necessarily maintained when $\eta_a >0$ and it is sensitive to the nature of the imposed metric stimulus. Panels (a), (b) and (c) in fig.~\ref{FIG:RingLikeA} show the shape accommodated for $\eta_a=0.5$, $\eta_a=1$ and $\eta_a=2$, respectively. The dome-like morphology is maintained for an isotropic metric stimulus, $\eta_a=1$, while the onset of a more protruded or a ring-like morphology can be seen for $\eta_a=0.5$ and $\eta_a=2$, respectively. Although those three shapes are qualitatively similar to the ones obtained when imposing only a curvature stimulus, the strain profiles are dramatically different as shown in panels (d), (e) and (f) in fig.~\ref{FIG:RingLikeA}. Compressive azimuthal strains are now located within the stimulated domain; the meridional component of the strain in the stimulated domain remains tensile, but a much lower magnitude, or turns even compressive as in the case with $\eta_a=2$.

\section{Conclusion}

In this work, we derive two reduced two-dimensional energy functionals to describe the nonlinear mechanics of naturally curved morphoelastic shells made of an incompressible Neo-Hookean material and a compressible Ciarlet–Geymonat material. Starting from the energies described in three dimensions, the reduced forms are obtained performing an asymptotic expansion exploiting the smallness of the shell thickness compared to the in-plane characteristic length. 

The formulation proposed retains geometric and constitutive nonlinearities without imposing additional kinematic constraints. In doing so, we describe the through-the-thickness behavior following the approach presented by Ozenda and Virga \cite{Ozenda2021} and extending it to the case of a naturally curved shell. Furthermore, we embed the non-elastic morphing stimuli by decomposing them into (i) in-plane metric and curvature components and (ii) through-the-thickness component, thus enabling their independent control. The terms of order $O(h_0^3)$ in the energies correctly retain all the contributions that must consistently appear at this order, including the quadratic contributions in $Z$ in the form of the functions $\phi$ and $\bar{\phi}$ as well as in the contravariant metric $\Gci_{\alpha \beta}$. In the energy functionals derived here, we assume constraints on the form $\cci_{\alpha \beta}$ in the expansion of the  metric $\Gci_{\alpha \beta}$: those constraints can be relaxed in future work to make us able to investigate the behavior under a arbitrary intermediate metric.

We validated the reduced model against two representative problems already discussed in the literature: the eversion of spherical caps and the eversion of a closed sphere. The study of those examples illustrates the importance of the appropriate modeling of the through-the-thickness response and the need to retain the coupling between geometry and constitutive behavior, also in the limit of small strains and moderate rotations. Finally, we explored how the framework can be applied to morphoelastic strategies relevant for budding and vesiculation, pointing to possible future directions for studying localized shape changes in active and biological solid-like shells. 
\vskip2pc

{\bf Acknowledgement.} M.T. is a member of the Gruppo Nazionale di Fisica Matematica (GNFM) of the Istituto Nazionale di Alta Matematica (INdAM). We thank Souhayl Sadik for insightful discussions about this work.

\vskip6pt
\vskip2pc

\clearpage
\appendix

\section{Expansion of the invariants}
In this Section, we present further details of results that are used in the main text. We recall that we consider a body $\mathcal{B}_0 =\mathcal{S}_0 \times [-h_0/2,h_0/2]$, where $\mathcal{S}_0$ defines the 2D midsurface of the object and $h_0$ is its thickness. Its 3D covariant metric  $\Gcr_{ij}$, with $i,j=1,2,3$, with respect to a set of a curvilinear material coordinate system $[X^1, X^2, X^3]$, in the reference configuration takes the form
\begin{equation}
    \Gcr_{ij} \upd X^i \upd X^j = \Gcr_{\alpha \beta} \upd \eta^{\alpha} \upd \eta^{\beta}+Z^2 \upd Z \upd Z\,,
\end{equation}
where the metric is decoupled in the in-plane contribution, denoted by $\Gcr_{\alpha \beta}$, with the Greek indices $\alpha,\beta = 1,2$, and the through-the-thickness contribution. We also have introduced a set of curvilinear coordinates $[\eta^{1}, \eta^2, Z]$, where $\eta^\alpha$ (with $\alpha=1,2$) identify the in-plane coordinates, while $Z$ is the coordinate along the normal direction. As discussed in the main text, the components of the metric associated with these two contributions take the form 
\begin{align}
    \Gcr_{\alpha \beta} = \acr_{\alpha \beta}(\eta^{\alpha}) - 2 Z \bcr_{\alpha \beta} (\eta^{\alpha}) + Z^2 \ccr_{\alpha \beta}(\eta^{\alpha}), \qquad \Gci_{33} = 1\,,
\end{align}
where $\acr_{\alpha \beta}$, $\bcr_{\alpha \beta}$, and $\ccr_{\alpha \beta}$ are the first, second, and third fundamental forms of the mid-surface $\mathcal{S}_0$, respectively.
By introducing two functions $\phi(\eta^{\alpha},Z)$ and $\iphi(\eta^{\alpha},Z)$ describing the through-the-thickness behavior in the current and intermediate configuration, respectively, we can also write the components of the covariant metrics in these two configurations as
\begin{align}
    \Gcc_{\alpha \beta} &= \acc_{\alpha \beta}(\eta^{\alpha}) - 2\phi(\eta^{\alpha},Z) \bcc_{\alpha \beta}(\eta^{\alpha}) + \phi^2(\eta^{\alpha},Z) \ccc_{\alpha \beta}(\eta^{\alpha})\,, &\Gcc_{33} = \left(\phi'\right)^2\,,
\end{align}
and
\begin{align}
    \Gci_{\alpha \beta} &= \aci_{\alpha \beta}(\eta^{\alpha}) - 2\iphi(\eta^{\alpha},Z)\bci_{\alpha \beta}(\eta^{\alpha}) + \iphi^2(\eta^{\alpha},Z) \bar{c}_{\alpha \beta}(\eta^{\alpha})\,, &\Gci_{33} = \left(\iphi'\right)^2\,.
\end{align}
We notice that $\acc_{\alpha \beta}$, $\bcc_{\alpha \beta}$, and $\ccc_{\alpha \beta}$ are the components of the first, second, and third fundamental forms of the mid-surface in the current configuration, while $\aci_{\alpha \beta}$, $\bci_{\alpha \beta}$, and $\cci_{\alpha \beta}$ are the components of the target first, second, and third forms of the mid-surface in the intermediate configuration.
 
\subsection{Third invariant}\label{SM:Third invariant}
In the main text, we rewrite the third invariant as 
\begin{align}
    I_3 = \frac{\det \left(\Gcc_{ij}\right)}{\det\left(\Gci_{ij}\right)} = \left(\frac{\phi'}{\iphi'}\right)^2\frac{\det \left(\Gcc_{\alpha \beta}\right)}{\det\left(\Gci_{\alpha \beta}\right)}\,.
\end{align}
To derive the expansions of $\det \left(\Gcc_{\alpha \beta}\right)$ and $\det \left(\Gci_{\alpha \beta}\right)$ in the functions $\phi$ and $\iphi$, we treat $\Gc_{\alpha \beta}$ and $\Gi_{\alpha \beta}$ as tensorial quantities. For simplicity of notation, we show the passages on the metric $\Gcr_{\alpha \beta}$ expanded in the more familiar coordinate $Z$ but the same process applies for the metrics in all configurations.

We know that, for generic rank 2 tensors ${\bf A}$ and ${\bf B}$, it is true that $\det\left({\bf A}+{\bf B}\right)=\det\left({\bf A}\right) + \det \left({\bf B}\right) + \det\left({\bf A}\right) \mathrm{tr}\left({\bf A}^{-1}{\bf B}\right)$. Employing the linearity of the trace operator we obtain the following series of equalities
\begin{align}
    \notag \det{\Gcr_{\alpha\beta}}& = \det(\acr_{\alpha\beta}) + \det(- 2 Z \bcr_{\alpha \beta} + Z^2 \ccr_{\alpha \beta}) - 2 Z \det(\acr_{\alpha\beta}) \acr^{\alpha\beta}\bcr_{\alpha\beta}  + Z^2 \det(\acr_{\alpha\beta}) \acr^{\alpha\beta}\ccr_{\alpha\beta}\\
    \notag &= \det(\acr_{\alpha\beta})  +4 Z^2 \det(\bcr_{\alpha \beta}) -2 Z \det(\acr_{\alpha\beta}) \acr^{\alpha\beta}\bcr_{\alpha\beta}  + Z^2 \det(\acr_{\alpha\beta}) \acr^{\alpha\beta}\ccr_{\alpha\beta} + O(Z^3)\\
    \notag &=\det(\acr_{\alpha\beta})\left[1 -2 Z \acr^{\alpha\beta}\bcr_{\alpha\beta} +Z^2 \left(4\frac{\det(\bcr_{\alpha \beta})}{\det(\acr_{\alpha\beta})}+ \acr^{\alpha\beta}\ccr_{\alpha\beta}\right) \right] + O(Z^3)\\
    &=\det(\acr_{\alpha\beta})\left[1 -4 Z\overset{\circ}{H}  +Z^2 \left(2\overset{\circ}{K}+  4\overset{\circ}{H}^2\right) \right] + O(Z^3)\,,
\end{align}
where we used the relations between fundamental forms and curvatures as $\acr^{\alpha \beta} \bcr_{\alpha \beta} = 2 \overset{\circ}{H}$, $\det\left(\bcr_{\alpha \beta}\right)/\det\left(\acr_{\alpha\beta}\right) = \overset{\circ}{K}$ and $\acr^{\alpha \beta} \ccr_{\alpha \beta} = 4\overset{\circ}{H}^2 -2 \overset{\circ}{K}$.

\subsection{Second invariant} \label{SM:Second invariant}
In deriving the second invariant $I_2$ of the right Cauchy-Green tensor ${\bf C}_e$, written as
\begin{align}\label{eq:Ce}
    {\bf C}_e = \Gcc_{ij} \left(\gi^i \otimes \gi^j\right),
\end{align}
where $\gi^i$ indicates the tangent vector in the intermediate configuration, we want to show that 
\begin{equation}
    I_2= \frac{1}{2} \left(\mathrm{tr}^2({\bf C}_e)-\mathrm{tr}({\bf C}^2_e)\right)\,,
\end{equation}
is equivalent to 
\begin{equation}
    I_2 = I_3 \left(\frac{\phi'}{\iphi'}\right)^{-2} +\left(\frac{\phi'}{\iphi'}\right)^2 \left(I_1-\left(\frac{\phi'}{\iphi'}\right)^2\right)\,,
\end{equation}
where $I_1$ is the first invariant of ${\bf C}_e$, expressed as
\begin{align}\label{eq:I1}
    I_1 &= \Gcc_{\alpha \beta} \Gci^{\alpha \beta} + \Gcc_{33} \Gci^{33} =  \Gcc_{\alpha \beta} \Gci^{\alpha \beta}+\left(\frac{\phi'}{\iphi'}\right)^2\,,
\end{align}
and
\begin{align}\label{eq:I3}
    I_3 = \left(\frac{\phi'}{\iphi'}\right)^2\frac{\det \left(\Gcc_{\alpha \beta}\right)}{\det\left(\Gci_{\alpha \beta}\right)}\,.
\end{align}

By inserting the form of the covariant current metric, as expressed in eq.~\eqref{eq:CovariantMetricCurrent}, in eq.~\eqref{eq:Ce}, the right Cauchy-Green tensor becomes 
\begin{equation}
    {\bf C}_e=\Gcc_{\alpha \beta} \left(\gi^{\alpha} \otimes \gi^{\beta}\right) + \left(\phi'\right)^2 \left(\gi^3 \otimes \gi^3\right)\,.
\end{equation}
We redefine  the identity tensor $ {\bf I} = \gi_i \otimes \gi^i$ via the two projection operators as $\mathbf{I}=\mathbf{N}+\mathbf{P}$ defined as and $\mathbf{P} = \gi_\alpha \otimes \gi^\alpha $ and $\mathbf{N} =\gi_3 \otimes \gi^3$. We first notice that
\begin{equation}
    {\bf I}: \left(\gi^3 \otimes \gi^3\right) =\left(\gi_l \otimes \gi^l\right):\left(\gi^3 \otimes \gi^3\right) = \delta_l^3\left(\gi^l \cdot \gi^3\right) = \Gci^{33} = \left(\iphi'\right)^{-2}\,,
\end{equation}
and
\begin{equation}
    \Gcc_{\alpha \beta} \left(\gi^{\alpha} \otimes \gi^{\beta}\right) \left(\gi^3 \otimes \gi^3\right)= \Gcc_{\alpha \beta}\left(\gi^{\beta} \cdot \gi^3\right)\left(\gi^{\alpha} \otimes \gi^3\right) = \bf{0}\,.
\end{equation}
We can thus write
\begin{align}
    &\left({\bf I}:{\bf C}_e\right)^2 = \left({\bf P}:\Gcc_{\alpha \beta} \left(\gi^{\alpha} \otimes \gi^{\beta}\right)\right)^2 + 2 \left(\frac{\phi'}{\iphi'}\right)^2 \left({\bf P}:\Gcc_{\alpha \beta} \left(\gi^{\alpha} \otimes \gi^{\beta}\right)\right)  +\left(\frac{\phi'}{\iphi'}\right)^4,\\
    &{\bf I}:{\bf C}_e^2 ={\bf P}: \left(\Gcc_{\alpha \beta} \left(\gi^{\alpha} \otimes \gi^{\beta}\right)\right)^2 +\left(\frac{\phi'}{\iphi'}\right)^4.
\end{align}
This allows us to rewrite the invariant as 
\begin{equation}\label{eq:I2_intermediate}
    I_2 = \frac{1}{2} \left(\left({\bf P}:\Gcc_{\alpha \beta} \left(\gi^{\alpha} \otimes \gi^{\beta}\right)\right)^2-{\bf P}:\left(\Gcc_{\alpha \beta} \left(\gi^{\alpha} \otimes \gi^{\beta}\right)\right)^2\right) +\left(\frac{\phi'}{\iphi'}\right)^2 \left({\bf P}:\Gcc_{\alpha \beta} \left(\gi^{\alpha} \otimes \gi^{\beta}\right)\right) \,.
\end{equation}
Applying the Cayley-Hamilton theorem to the term $\Gcc_{\alpha \beta} \left(\gi^{\alpha} \otimes \gi^{\beta}\right)$, we can write
\begin{equation}
    \Gcc_{\alpha \beta} \left(\gi^{\alpha} \otimes \gi^{\beta}\right)-\left({\bf P
    }:\Gcc_{\alpha \beta} \left(\gi^{\alpha} \otimes \gi^{\beta}\right)\right)\Gcc_{\alpha \beta} \left(\gi^{\alpha} \otimes \gi^{\beta}\right) + \mathrm{det}\left(\Gcc_{\alpha \beta} \left(\gi^{\alpha} \otimes \gi^{\beta}\right)\right){\bf P} = {\bf 0}\,.
\end{equation}
Moreover, taking the (in-plane) trace with respect to $\mathbf{P}$ on both sides, we obtain
\begin{equation}
    {\bf P}: \Gcc_{\alpha \beta} \left(\gi^{\alpha} \otimes \gi^{\beta}\right) - \left({\bf P
    }:\Gcc_{\alpha \beta} \left(\gi^{\alpha} \otimes \gi^{\beta}\right)\right)^2 + 2 \mathrm{det}\left(\Gcc_{\alpha \beta} \left(\gi^{\alpha} \otimes \gi^{\beta}\right)\right) = 0\,,
\end{equation}
which we can thus rewrite as
\begin{equation}
    \left({\bf P
    }:\Gcc_{\alpha \beta} \left(\gi^{\alpha} \otimes \gi^{\beta}\right)\right)^2 -{\bf P}: \Gcc_{\alpha \beta} \left(\gi^{\alpha} \otimes \gi^{\beta}\right) = 2 \mathrm{det}\left(\Gcc_{\alpha \beta} \left(\gi^{\alpha} \otimes \gi^{\beta}\right)\right)\,.
\end{equation}
Finally, substituting this expression in eq.~\eqref{eq:I2_intermediate}, the second invariant can be written as
\begin{align}
    \notag I_2 &= \mathrm{det}\left(\Gcc_{\alpha \beta} \left(\gi^{\alpha} \otimes \gi^{\beta}\right)\right) +\left(\frac{\phi'}{\iphi'}\right)^2 \left({\bf P}:\Gcc_{\alpha \beta} \left(\gi^{\alpha} \otimes \gi^{\beta}\right)\right) = \frac{\det \left(\Gcc_{\alpha \beta}\right)}{\det\left(\Gci_{\alpha \beta}\right)}+\Gcc_{\alpha \beta} \Gci^{\alpha \beta} \\
    &=I_3 \left(\frac{\phi'}{\iphi'}\right)^{-2}+\left(\frac{\phi'}{\iphi'}\right)^2 \left(I_1-\left(\frac{\phi'}{\iphi'}\right)^2\right)\,,
\end{align}
where, in the latter equality, we used eq.~\eqref{eq:I1} and eq.~\eqref{eq:I3}.

\section{Full form of the reduced energies}
In this sections we present the full form of the reduced energies, when the through-the-thickness non-elastic stimulus includes also $\bar{\rho}_2\neq0$ and $\bar{\rho}_3\neq0$, other than the contribution of $\bar{\rho}_1$.
\subsection{Neo-Hookean material}\label{SM:NH_fullform}
As shown in the main document, the energy functional of the effective two-dimensional Neo-Hookean shell becomes
\begin{align}
    \Psi_{nc,2d} = \int_{\mathcal{S}_0} \left[ h_0 w^\textup{NH}_s + \frac{h_0^3}{12}w^\textup{NH}_b \right] \sqrt{\det(\aci_{\alpha \beta})} \, \upd \eta^1 \upd \eta^2 + O(h^5)\,,
\end{align}
where the integral is evaluated with respect to the coordinates in the reference configuration $\mathcal{S}_0$ and $h_0$ indicates the undeformed thickness. The stretching contribution $w^\textup{NH}_s$ depends only on $\bar{\rho}_1$ so it remains equal to what has been presented in the main text, while the bending term in the iso-areal case is modified by $\bar{\rho}_1$ and $\bar{\rho}_2$ as
\begin{align}
    \notag w^\textup{NH}_b = & \frac{\mu^*}{2}\left( 2\frac{\bar{K}}{\bar{\rho}_1}\left(\bar{\rho}_1^3 \bar{K}-6 \bar{\rho}_1 \bar{\rho}_2 \bar{H}+3 \bar{\rho}_3\right)  w^\textup{NH}_s+ 2 \bar{\rho}_1 \left(\bar{\rho}_1^2 (H+\bar{H})-3 \bar{\rho}_2\right)\left(H-\bar{H}\right) \bcc_{\alpha \beta} \aci^{\alpha \beta} \right. \\
    & \left.+2 \bar{\rho}_1 \left(3 \bar{\rho}_2-2 \bar{\rho}_1^2 \bar{H}\right) \bar{H} \acc_{\alpha \beta} \bci^{\alpha \beta} +  p_0 f_{p0}+p_1 f_{p1}\right.\\
    &\left.+  \bar{\rho}_1^3\left( - 4 \bcc_{\alpha \beta} \bci^{\alpha \beta} + \ccc_{\alpha \beta} \aci^{\alpha \beta} +  3\acc_{\alpha \beta} \cci^{\alpha \beta}\right) + f_1\right)\,,
\end{align}
with
\begin{align}
    f_1 &= 2\bar{\rho}_1 \left(\bar{\rho}_1^2 (2 (H-\bar{H}) (4 H+\bar{H})-K+\bar{K})+6 \bar{\rho}_2 (\bar{H}-H)\right) ,\\
    f_{p0} &=\frac{24}{\mu^*}   \bar{\rho}_1 J_s (H-\bar{H}) \left(\bar{\rho}_1^2 H-\bar{\rho}_2\right)\,\\
    f_{p1} &=\frac{4}{\mu^*} \bar{\rho}_1^2 (-2 H J_s+\bar{H} J_s+\bar{H})+\bar{\rho}_2 (J_s-1)\,.
\end{align}

\subsection{Ciarlet-Geymonat material}\label{SM:CG_fullform}
Using the same notation as in the main document, the energy functional of the effective two-dimensional Ciarlet-Geymonat shell becomes, when rescaled by $\mu^*$,
\begin{align}
    \left(\mu^*\right)^{-1}\Psi_{cg} = \int_{\mathcal{S}_0} \left[ h_0 w^{CG}_s + \frac{h_0^3}{12} w^{CG}_b \right] \sqrt{\det(\aci_{\alpha \beta})} \, \upd \eta^1 \upd \eta^2 + O(h^5)\,,
\end{align}
where we define the rescaled bulk modulus as $\lambda = \lambda^*/\mu^*$. The stretching contribution $ w^\textup{CG}_s$ has been already presented in the main text because it does not depend on the higher order terms in the through-the-thickness, $\bar{\rho}_2$ and $\bar{\rho}_3$. The bending contribution is instead modified as
\begin{align}\label{eq:CG_BendingTOT}
    w_b^\textup{CG} &= g_1 w^\textup{CG}_s +g_2 \bcc_{\alpha \beta} \aci^{\alpha \beta} +g_3 \acc_{\alpha \beta} \bci^{\alpha \beta} - 2 \rho_1 \bar{\rho}_1^2 \bcc_{\alpha \beta} \bci^{\alpha \beta} + \frac{1}{2} \rho_1^2 \bar{\rho}_1 \ccc_{\alpha \beta} \aci^{\alpha \beta} - \frac{3}{2} \bar{\rho}_1^3 \acc_{\alpha \beta} \cci^{\alpha \beta} + g_4\,.
\end{align}
where 
\begin{align}
    g_1&=  \bar{\rho}_1^3  \bar{K}-6  \bar{\rho}_1  \bar{\rho}_2  \bar{H}+3  \bar{\rho}_3\,,\\
    g_2&=  \left(2 \rho_1  \bar{\rho}_1^2 \bar{H}-2 \rho_1  \bar{\rho}_2- \bar{\rho}_1 \rho_2\right)\,,\\
    g_3&= \bar{\rho}_1 \left(3 \bar{\rho}_2-2 \bar{\rho}_1^2 \bar{H}\right)\,,\\
     \nonumber g_4&= 
    \bar{\rho}_1 \left(\frac{\left(\rho_2^2-5\rho_1^2\bar{\rho}_2\bar{H}\right) (\lambda+2)}{\rho_1^2}+\frac{1}{2} \rho_1^2 \left(2 H^2 (\lambda+2 )+2 \bar{H}^2 J_s \lambda-\lambda (J_s \bar{K}+K)-2 K\right)+ \rho_2 H (\lambda+2)\right)\\
   \nonumber &+\bar{\rho}_1^2\frac{2 \bar{H}(\lambda+2) \left(\rho_2-\rho_1^2 H\right)}{\rho_1}+\frac{1}{2} \bar{\rho}_1^3 \left(2 \bar{H}^2+\bar{K}\right) (\lambda+2)\\
   \nonumber &+\frac{\bar{\rho}_1^{-1}}{2}\left(\rho_1^4 J_s K \lambda+4  \left(\bar{\rho}_2 \left(2 \rho_1^2 \bar{H}+\bar{\rho}_2\right)+\rho_2^2\right)+2 l \left(J_s \left(\rho_1^4 H^2-5 \rho_1^2 \rho_2 H+\rho_1^2 \bar{\rho}_2 \bar{H}+\rho_2^2\right)+\bar{\rho}_2^2\right)\right)\\
    \nonumber&+\frac{\bar{\rho}_1^{-2}\rho_1}{2} \left(\rho_1 (4 \rho_1 \bar{\rho}_2 H J_s\lambda-3 \bar{\rho}_3 (J_s \lambda+2 ))-4 \rho_2 \bar{\rho}_2  (J_s \lambda+2 )\right)\\
    \nonumber&+\bar{\rho}_1^{-3}\left(\rho_1^2 \bar{\rho}_2^2 \left(J_s \lambda+2\right)\right)\\
    &+\bar{\rho}_2 \left(2 \rho_1  H (\lambda+2 )-\frac{2 \rho_2 (\lambda+2 )}{\rho_1}\right)+\frac{3}{2} \bar{\rho}_3 (\lambda+2 )-2 \rho_1 \bar{H} \left(\rho_1^2 H J_s \lambda-\rho_2 J_s \lambda+2 \rho_2\right)\,.
\end{align}
Minimization of eq. \eqref{eq:CG_BendingTOT} with respect to $\rho_2$ leads to:
\begin{align}
    \rho_2 = \frac{1}{2} \bar{\rho}_1^2 \left[\left(\frac{H(\lambda + 2)(2 J_s \lambda - 1)}{(J_s \lambda + 2)^2} - \frac{2 \bar{H} J_s \lambda \sqrt{\lambda + 2}}{(J_s \lambda + 2)^{3/2}} \right) + \frac{\bcc_{\alpha \beta} \aci^{\alpha \beta}}{4 J_s \lambda + 8}\right]+\bar{\rho}_2\frac{\sqrt{\lambda+2}}{\sqrt{J_s \lambda+2 }}\,,
\end{align}
with $\rho_1$ as presented in the main text.
\bibliographystyle{abbrv}
\bibliography{sample_arxiv}

\end{document}